\documentclass[review, preprint, 12pt]{elsarticle}

\usepackage{amsmath}
\usepackage{url}
\usepackage{float}
\usepackage{subcaption}
\usepackage{longtable}
\usepackage{algorithm}
\usepackage{algpseudocode}

\begin{document}

\begin{frontmatter}

\title{Predicting cement microstructure and mechanical properties in hydrating cement paste with a Phase-Field model}

\author[1,2,3]{Alexandre Sac-Morane\corref{cor1}}
\ead{alexandre.sac-morane@enpc.fr}

\author[4]{Katerina Ioannidou}
\ead{aikaterini.ioannidou@umontpellier.fr}

\author[1]{Manolis Veveakis}
\ead{manolis.veveakis@duke.edu}

\author[2]{Hadrien Rattez}
\ead{hadrien.rattez@uclouvain.be}

\cortext[cor1]{Corresponding author}

\affiliation[1]{organization={Multiphysics Geomechanics Lab, Duke University}, addressline={Hudson Hall Annex, Room No. 053A}, city={Durham}, postcode={27708}, country={NC, USA}}

\affiliation[2]{organization={Institute of Mechanics, Materials and Civil Engineering, UCLouvain}, addressline={Place du Levant 1}, city={Louvain-la-Neuve}, postcode={1348}, country={Belgium}}

\affiliation[3]{organization={Laboratoire Navier, ENPC, Institut Polytechnique de Paris, Univ Gustave Eiffel}, addressline={12 boulevard Copernic}, city={Champs-sur-Marne}, postcode={77420}, country={France}}

\affiliation[4]{organization={Laboratory of Mechanics and Civil Engineering, University of Montpellier}, addressline={860 Rue de St-Priest}, city={Montpellier}, postcode={34090}, country={France}}

\begin{abstract}
Predicting the evolving microstructure of hydrating cement is essential for understanding and modeling its mechanical property development. Physics-based continuum approaches offer a rigorous framework for capturing the thermodynamics of dissolution and precipitation processes at the microstructural scale. In this work, we present an adapted Phase-Field (PF) model for cement hydration that resolves key physical inconsistencies in existing PF formulations by introducing a revised free-energy potential and distinct equilibrium constants for clinker dissolution and hydrate precipitation. The resulting PF framework reproduces microstructural evolution, yielding realistic porosity levels and continuous phase boundaries in close agreement with experimental observations.

The predicted hydrated microstructures are subsequently used in a computational homogenization scheme to evaluate the elastic response of the material. The PF-derived mechanical properties show good agreement with experimental trends, supporting the ability of the proposed framework to consistently link hydration chemistry, microstructure formation, and the resulting mechanical response.
\end{abstract}

\begin{highlights}
    \item A physics-based Phase-Field model predicts the microstructure evolution during the hydration of cement-based material.
    \item Predicted microstructures are used to derive the elastic moduli of the composite material with a computational homogenization scheme.
    \item Predicted microstructures and elastic moduli show excellent agreement with data available in the literature. 
\end{highlights} 


\end{frontmatter}


\section{Introduction}

A central objective in the study of cement-based materials is to elucidate and predict their mechanical behavior \cite{Nithurshan2023}, in order to enable the optimization of existing formulations or the development of alternative binders \cite{Shah2022}, particularly those with reduced environmental impact. Over the past decades, numerous analytical and empirical models have been proposed to capture the wide range of factors governing the mechanical performance of cementitious systems \cite{Feret1892, Chidiac2013, Nithurshan2023}. These factors include, among others, the water-to-cement mass ratio ($w/c$), raw material properties, particle size distribution, and the presence of chemical or mineral additives \cite{DeLarrard1999, Thomas2009, Kawashima2013, Lefever2020, Nguyen2021}.

Beyond these empirical formulations, foundational physical models were developed to describe the evolution of the hydrate microstructure. The Powers–Brownyard model \cite{Powers1946} provided the first quantitative framework for the structure of hydrated cement paste, distinguishing between gel pores and capillary pores. Later, idealized microstructures were described using homogenization approaches, involving three steps—description, localization, and upscaling—to relate microscale phase morphology to macroscale mechanical properties \cite{Agofack2015, Samudio2017}. A variety of homogenization schemes have been proposed \cite{Budiansky1965, McLaughlin1977, Ghabezloo2010, Halpin1976, Pichler2009}, with the dilute \cite{Eshelby1957}, self-consistent \cite{Hill1965}, and Mori–Tanaka \cite{Mori1973} approaches being the most widely adopted. Such schemes can be concatenated across multiple scales—from hydrate foam (20 µm) to cement paste (0.7 mm), and from cement paste to mortar (1 cm) \cite{Pichler2011, Pichler2013, Pichler2008}. Coupling these approaches with hydration kinetics enables the prediction of evolving mechanical properties throughout hydration \cite{Wang2017}. These efforts established the conceptual link between hydration degree, porosity, and mechanical performance—a relationship modern computational models now aim to simulate explicitly \cite{Bentz2006b}. Nonetheless, these homogenization methods rely on idealized or simplified microstructures.

To overcome such limitations, microstructure-explicit hydration models were developed. One of the earliest is HYMOSTRUC \cite{vanBreugel1995}, where reactants and products are idealized as continuum spheres whose sizes evolve during dissolution or precipitation. This framework was later extended to voxel-based microstructures in CEMHYD3D \cite{Holmes2020}, where hydration is simulated through Cellular Automata (CA). Although widely used, CA approaches are constrained by voxel resolution. To alleviate this limitation, alternative solvers such as HydratiCA \cite{Bullard2007}, $\mu$ic \cite{Bishnoi2009}, and the Integrated Particle Kinetics Model \cite{Pignat2005} were introduced. However, all those methods are based on semi-empirical reaction coefficients. 

More recently, level-set based approaches were proposed to describe microstructure evolution using continuous fields and physics-based dissolution–precipitation laws \cite{NguyenTuan2024}. Although these methods incorporate fundamental hydration kinetics, the underlying Kooi law \cite{Kooi2006} is formulated at the macroscale and does not strictly satisfy local physical constraints at the microstructural scale.

To address this issue, Phase-Field (PF) methods have emerged as a promising alternative \cite{Petersen2018}. PF formulations naturally describe interface evolution based on local thermodynamic principles and have been successfully applied to various dissolution/precipitation \cite{Moelans2008, Takaki2014, Guevel2020, Li2023, SacMorane:PFDEM, SacMorane:PFDEMb} and fracture problems \cite{Ulloa2021, Xing2023, Wu2025}. When coupled with diffusion equations for dilute species, PF models enable dissolution and precipitation to occur whenever local solute concentrations deviate from their equilibrium values, thus ensuring thermodynamic consistency at the microstructural scale.

However, the formulation introduced in \cite{Petersen2018} presents two limitations that motivate the present work. First, the free-energy functional employed permits spontaneous gel precipitation even when the system is at equilibrium. Second, the equilibrium concentrations associated with source–solute (dissolution) and solute–gel (precipitation) reactions are assumed equal, whereas they are known to be distinct according to established hydration theory \cite{Bullard2011, Scrivener2015}. These inconsistencies hinder the model’s physical fidelity and predictive capability. To address these issues, the present work proposes an adapted Phase-Field formulation in which the free-energy landscape and equilibrium relationships are revised to better reflect the thermodynamics of cement hydration.

Once the hydrated microstructure has been computed, the mechanical behavior of the hardened cement-based material can be determined using computational homogenization. This permits the calculation of properties such as Young’s modulus, shear modulus, Poisson’s ratio, bulk modulus, and strength parameters, which depend explicitly on microstructural morphology and hydration state \cite{Guevel2022, Lindqwister2025}. The phases (pore, skeleton, binder) can be distinguished and characterized based on their intrinsic properties. The composite microstructure is then subjected to a Finite Element or Fast Fourier Transform-based homogenization scheme to infer macroscale mechanical parameters \cite{Isleem2023, Nguyen2024, Alavoine2020}.

The contribution of this work is organized as follows. First, the formulation of the adapted PF model for cement hydration is presented. The framework is then applied to boundary-value problems for $w/c=0.3$ and $0.5$. The microstructural evolution predicted by the model is compared with results from CEMHYD3D \cite{Nguyen2024} and with experimental measurements \cite{Thomas2009, Maruyama2014}. Finally, the resulting hydrated microstructures are subjected to mechanical loading (shear and tension) in order to estimate mechanical properties such as Young’s and shear moduli, which are compared to values reported in the literature, particularly those in \cite{Nguyen2024}.


\section{Formulation of the microstructure evolution}

\begin{figure}[ht]
    \centering
    \includegraphics[width=0.9\linewidth]{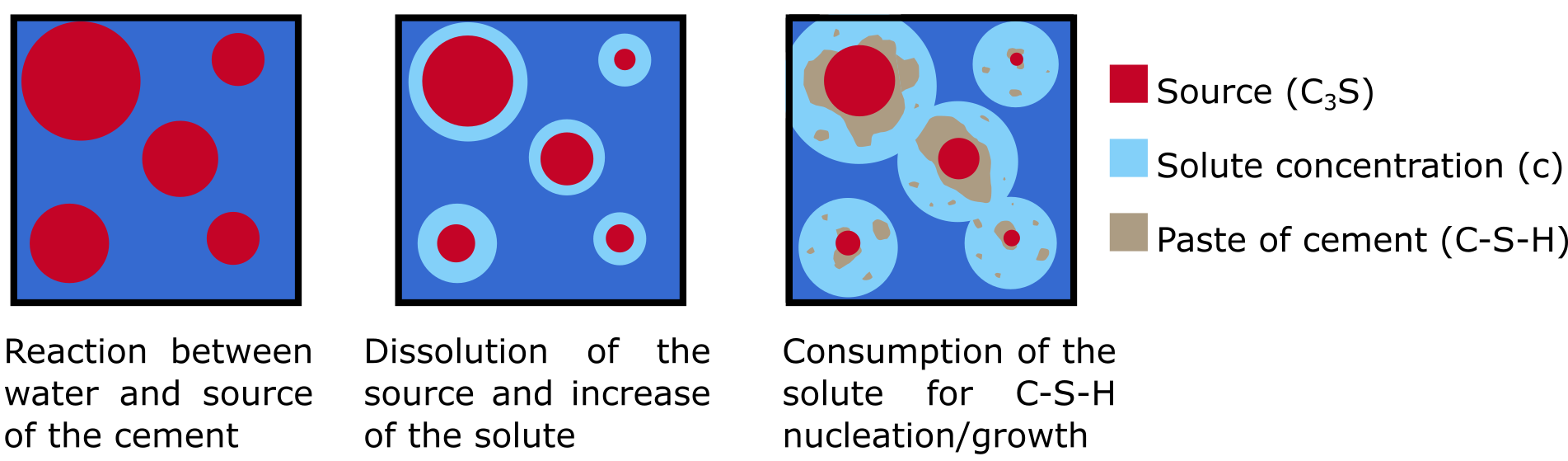}
    \caption{Scheme of the hydration process considered.}
    \label{Cementation Scheme Figure}
\end{figure}

This section focuses on the modelization of the microstructure evolution during hydration. This chemical process is divided into two main reactions: dissolution and precipitation. The source phases (mainly C$_3$S, C$_2$S, C$_3$A and C$_4$AF) react with water and dissolve \cite{Taylor1997,Zhang2020}. The generated solute diffuses and is consumed during the precipitation of the phases of the cement paste (mainly calcium silicate hydrates C-S-H and Portlandite CH) \cite{Taylor1997, Zhang2020}.
For simplicity reasons, only three phases are considered in this paper: the source (C$_3$S) which is the main clinker phase of Portland cement \cite{Zhang2020, Bullard2010}, the paste of the cement (C-S-H), which represents the main product of the hydration \cite{Nguyen2024, Zhang2020}, and the solute (c), which is a transient specie dissolved in the water. 
As shown in Figure \ref{Cementation Scheme Figure}, C$_3$S dissolves to generate c, and c is consumed to generate C-S-H after its diffusion. 
In the same vein of simplification, the local packing fraction of the C-S-H is considered constant and homogeneous in the precipitated phase. This assumption is a limitation compared to the observations showing that this property is heterogeneous and evolves with the hydration process \cite{Ioannidou2016, Ioannidou2014, Hu2014}.
These hypotheses are a common and effective first-order approximation for estimating the evolution of bulk elastic properties, which is the primary focus of this investigation. Of course, the framework presented herein can be extended to the other phases and to an evolving local packing fraction for the C-S-H.

In the following sections, the equations solved and the numerical model are presented. Then, a validation of the microstructure is performed by comparison with the literature \cite{Thomas2009, Nguyen2024, Maruyama2014}.


\subsection{Equations solved}
\label{Equations Section}

Three coupled equations are solved with a finite element solver MOOSE \cite{MOOSE, MOOSE_PF}. Equations \ref{Eq on psi} and \ref{Eq on phi} are Allen-Cahn formulations \cite{Allen1979} applied to the phase variables $C3S$ (the source of the cement) and $CSH$ (the cement paste) to model heterogeneous dissolution/precipitation phenomena \cite{SacMorane:PFDEM, SacMorane:PFDEMb}. The variable $C3S$ (resp. $CSH$) equals 1 if it is located inside a source of cement (resp. the C-S-H gel) and equals 0 elsewhere. 
In the following of this article, a variable $\eta$ can be used, representing the variables $C3S$ or $CSH$.
In parallel, Equation \ref{Eq on c} is a diffusive equation applied to the variable $c$. This variable represents the solute concentration in the water of the dilute species. 

\begin{align}
    \frac{\partial\, C3S}{\partial t}&=-L_{C3S} \frac{\partial f_{C3S} + Ed}{\partial\, C3S}+L_{C3S}\, \kappa_{C3S}\, \nabla^2 {C3S} \label{Eq on psi}\\
    \frac{\partial\, CSH}{\partial t}&=-L_{CSH} \frac{\partial f_{CSH} + Ep}{\partial\, {CSH}}+L_{CSH}\, \kappa_{CSH}\, \nabla^2 {CSH} \label{Eq on phi}\\
    \frac{\partial\, c}{\partial t}&= -\alpha_{CSH}\frac{\partial \,{CSH}}{\partial t} -\alpha_{C3S}\frac{\partial \,{C3S}}{\partial t} + \kappa_c\,\nabla^2c \label{Eq on c}
\end{align}
Where, $L_{\eta}$ is the effective mobility of the phase $\eta$ (impacting the kinetics of the relation, $=1$ here), $\kappa_{\eta}$ is the gradient coefficient of the phase $\eta$ (influencing the width of the phase interface), $f_{\eta}$ is the free energy of the phase $\eta$, $Ed/Ep$ are the tilting free energies. The formulation of the free energies $f_{\eta}/Ed/Ep$ is described in Section \ref{Free Energy Tilting}. Finally, $\kappa_c$ is the diffusivity of the solute concentration $c$, defined Equation \ref{Solute Diffusivity} \cite{Petersen2018}. 

\begin{equation}
    \kappa_c = \kappa_{c0}\left(1-C3S\right)\,\text{exp}(-\kappa_{cexp}\,CSH)
    \label{Solute Diffusivity}
\end{equation}
Where, the term $1-C3S$ is a penalty term to avoid diffusion of the solute in the source of cement. Similarly, the term exp$(-\kappa_{cexp}\, CSH)$ models the decrease of the diffusivity of the solute in the gel \cite{Mainguy2000, Atkinson1984}, where $\kappa_{cexp}$ is the penalty coefficient. A factor exp$(-\kappa_{cexp})=1/400$ is used \cite{Lu2019}.

Furthermore, the gradient coefficient $\kappa_{\eta}$ is related to the free energy $f_\eta$ described in Section \ref{Free Energy Tilting} \cite{Takaki2014, SacMorane:PFDEMb}.

\begin{equation}
    \kappa_{\eta}\propto W_{\eta}\cdot\delta^2 
    \label{Equation kappa}
\end{equation}
Usually, $\delta = 4\Delta - 10\Delta$ \cite{Moelans2008,Takaki2014}, where $\Delta$ is the mesh discretization. $W_\eta$ is the energy barrier of the free energy $f_\eta$, see Equation \ref{free energy Equation}. Even if a strong mesh dependency exists because of the PF definition, a method to cancel this impact is detailed in \cite{SacMorane:PFDEMb}.

The terms $\frac{\partial c}{\partial t} = -\alpha_{CSH} \frac{\partial CSH}{\partial t} -\alpha_{C3S} \frac{\partial C3S}{\partial t}$ of the Equation \ref{Eq on c} ensure the conservation of the total mass \cite{Petersen2018, SacMorane:PFDEM, SacMorane:PFDEMb}. The coefficients $\alpha_{CSH}$ and $\alpha_{C3S}$ represent conversion terms ($1\,\eta$ gives $\alpha_{\eta}\;c$).
These coefficients can be determined with the mass conservation between the initial $\cdot_i$ and the final $\cdot_f$ configurations at the saturation limit, see Equation \ref{Mass Conservation Relation}. The saturation limit is defined as the total consumption of the source particles and the entire saturation of the domain by the C-S-H gel. 

\begin{equation}
    \alpha_{C3S}\, C3S_i + \alpha_{CSH}\, CSH_i + c_i = \alpha_{C3S}\, C3S_f + \alpha_{CSH}\, CSH_f + c_f 
    \label{Mass Conservation Relation}
\end{equation}

With:
\begin{itemize}
    \item $C3S_i=1/2.344$, here the saturation is assumed at $\left(w/c\right)_{sat} = 0.42 \Longleftrightarrow \frac{1-C3S_i}{C3S_i}\frac{\rho_{H20}}{\rho_{C3S}}=0.42$, with $w$ the mass of water, $c$ the mass of the source, and $\rho_{H20}/\rho_{C3S}=1000/3200$ \cite{Petersen2018}.
    \item $CSH_i=0$, there is no C-S-H gel in the initial state (except some insignificant nucleation seeds).
    \item $c_i=c_{eq,\,CSH}$, the solute concentration is at equilibrium with the C-S-H gel in the initial state.
    \item $C3S_f=0$, there is no source of cement in the final state, ensuring the full hydration.
    \item $CSH_f=1$, the domain is fully saturated with the C-S-H gel in the final state, ensuring the full hydration.
    \item $c_f=c_{eq,\,C3S}$, the solute concentration is at equilibrium with the source of cement C3S in the final state, ensuring the full hydration.
\end{itemize}
By the nature of the problem, the equilibrium constants $c_{eq,\,C3S}$ and $c_{eq,\,CSH}$ must be different \cite{Bullard2011, Scrivener2015}. Furthermore, it appears that $c_{eq,\,C3S}$ is several orders of magnitude larger than $c_{eq,\,CSH}$ \cite{Zhang2020}.
Thus, it can be assumed that $c_{eq,\,C3S}=1$ and $c_{eq,\,CSH}=0$.
Injecting these assumptions into the Equation \ref{Mass Conservation Relation} and assuming $\alpha_{CSH}=3$, it gives $\alpha_{C3S}=9.376$.
It is important to notice that the value of the coefficient $\alpha_{CSH}$ has been found after a sensitivity analysis available in \ref{alpha_CSH sensitivity}. Indeed, this coefficient affects the conversion term $\alpha_{C3S}$ between the source $C3S$ and the solute $c$ during the dissolution. If the value is too small, the solute $c$ produced is not large enough to saturate the pore fluid, and the supersaturation state observed in experiments is not verified \cite{Bullard2011, Scrivener2015}.
Similarly, the consequence of a small $\alpha_{CSH}$ coefficient is the fact that the source particles are significantly dissolved before the precipitation of the paste, not in agreement with the experimental observations.


\subsection{Tilting the free energy $f_{\eta}$ to obtain dissolution/precipitation}
\label{Free Energy Tilting}

\begin{figure}[ht]
    \centering
    \includegraphics[width=0.6\linewidth]{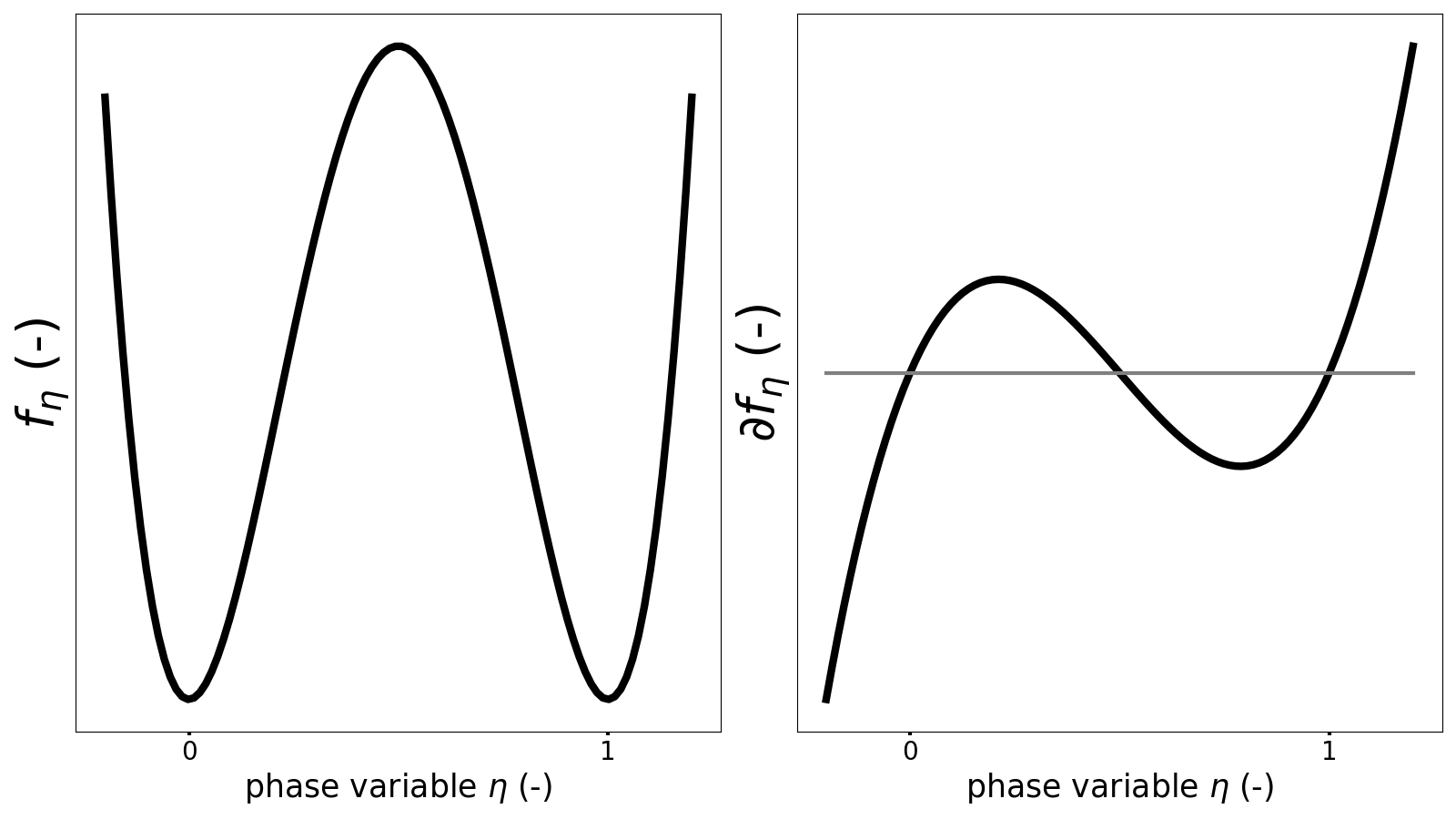}
    \caption{Free energy $f_{\eta}$ and its derivative $\partial f_{\eta}$.}
    \label{free energy Figure}
\end{figure}

In this description, the source of cement is modeled by the phase variable $C3S$. 
In a similar vein, the C-S-H gel is modeled by the phase variable $CSH$.
As a reminder, a variable $\eta$ can be employed herein to refer to $C3S$ or
$CSH$. 
As emphasized by the Allen-Cahn formulation detailed in Equations \ref{Eq on psi} and \ref{Eq on phi}, a free energy $f_{\eta}$ is employed to describe the problem. The equilibrium is reached in the system when the free energy is minimized. In the literature different expressions are available \cite{Petersen2018, Takaki2014, DeSousa2000, Chen2002}, but a double well one is preferred, as the minimum locations are fixed and known. Furthermore, it appears to be the most used form in the literature. The formulation of this function is given Equation \ref{free energy Equation} and is illustrated in Figure \ref{free energy Figure}.

\begin{equation}
    f_{\eta}({\eta}) = W_{\eta}\times{\eta}^2(1-{\eta})^2 
    \label{free energy Equation}
\end{equation}
Where, $W_{\eta}$ is the barrier energy of the phase variable ${\eta}$ ($=1$ here), preventing phase transformation.

\begin{figure}[ht]
    \centering
    \includegraphics[width=0.5\linewidth]{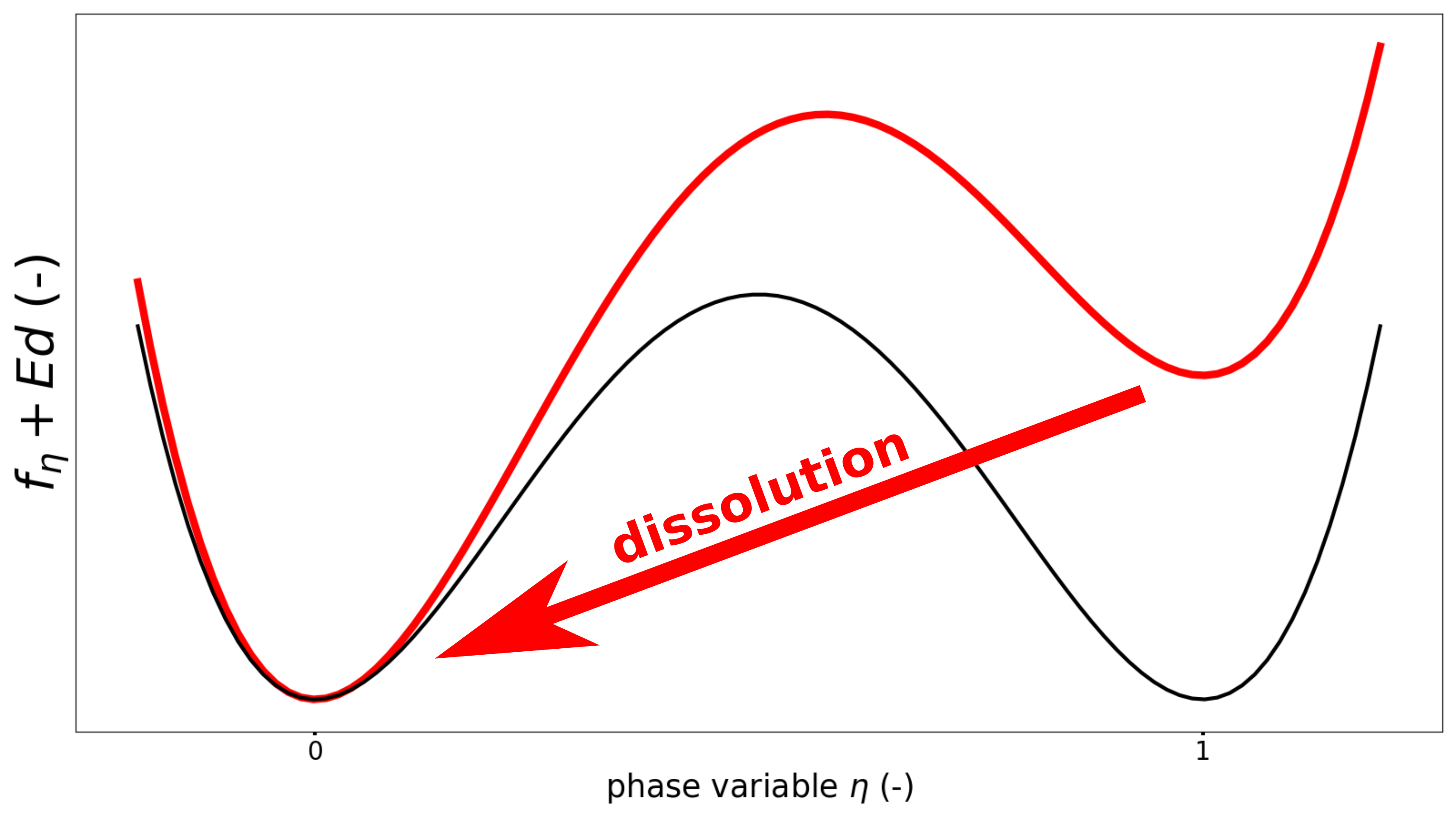}
    \caption{Example of tilted free energy $f_{C3S}+Ed(c, {C3S})$ (in red), the dissolution is favored ($c<c_{eq,\,C3S}$).}
    \label{tilted f_psi Figure}
\end{figure}

It is important to notice that the minima of this potential energy are located at ${\eta} = 0$ and ${\eta} = 1$. Then, these values are used to describe the source and the gel. As the PF description minimizes the free energy in the domain, the phase variable remains at these values at the equilibrium and without external destabilization. There is no dissolution/precipitation.
Subsequently, the idea is to add an external source term $Ed/Ep$ to tilt the initial free energy to favor localized dissolution or precipitation \cite{Petersen2018, Guevel2020, SacMorane:PFDEM, SacMorane:PFDEMb}. Indeed, depending on the tilting, ${\eta} = 0/1$ is no longer the minimum of the free energy, and some dissolution/precipitation occurs. These tilting free energy functions are defined Equations \ref{f_psi Ed Equation} and \ref{f_phi Ep Equation}.

\begin{equation}
    Ed(c, {C3S}) = \chi_{c\,,{C3S}}\times(c_{eq,\,C3S}-c)\times h_{C3S}({C3S})    
    \label{f_psi Ed Equation}
\end{equation}
Where, $h_{C3S}({C3S})=3\,{C3S}^2-2\,{C3S}^3$ is an interpolation function on the phase variable ${C3S}$ \cite{Takaki2014}, $\chi_{c,\,{C3S}}$ is the tilting coefficient related to the dissolution kinetics. This coefficient can be calibrated with experimental data. The choice of the interpolation function $h_{C3S}$ has an insignificant impact, while it verifies $h_{C3S}(0) = 0$, $h_{C3S}(1) = 1$, and $\frac{\partial h_{C3S}}{\partial C3S}(0) = \frac{\partial h_{C3S}}{\partial C3S}(1) = 0$.
For instance, other formulations have been employed in \cite{Petersen2018}.
It is important to notice the tilt occurs only if $c\ne c_{eq,\,C3S}$ and dissolution occurs in the case $c<c_{eq,\,C3S}$. The value $c_{eq,\,C3S}=1$ is considered for the equilibrium for the source dissolution. Figure \ref{tilted f_psi Figure} illustrates an example of tilting.
It is important to point out that the term $\chi_{c\,,{C3S}}\times(c_{eq,\,C3S}-c)$ determines the amplitude of the tilting, and so the kinetics of the dissolution.
The larger the tilting is, the faster the dissolution is.

\begin{figure}[ht]
    \centering
    \includegraphics[width=0.5\linewidth]{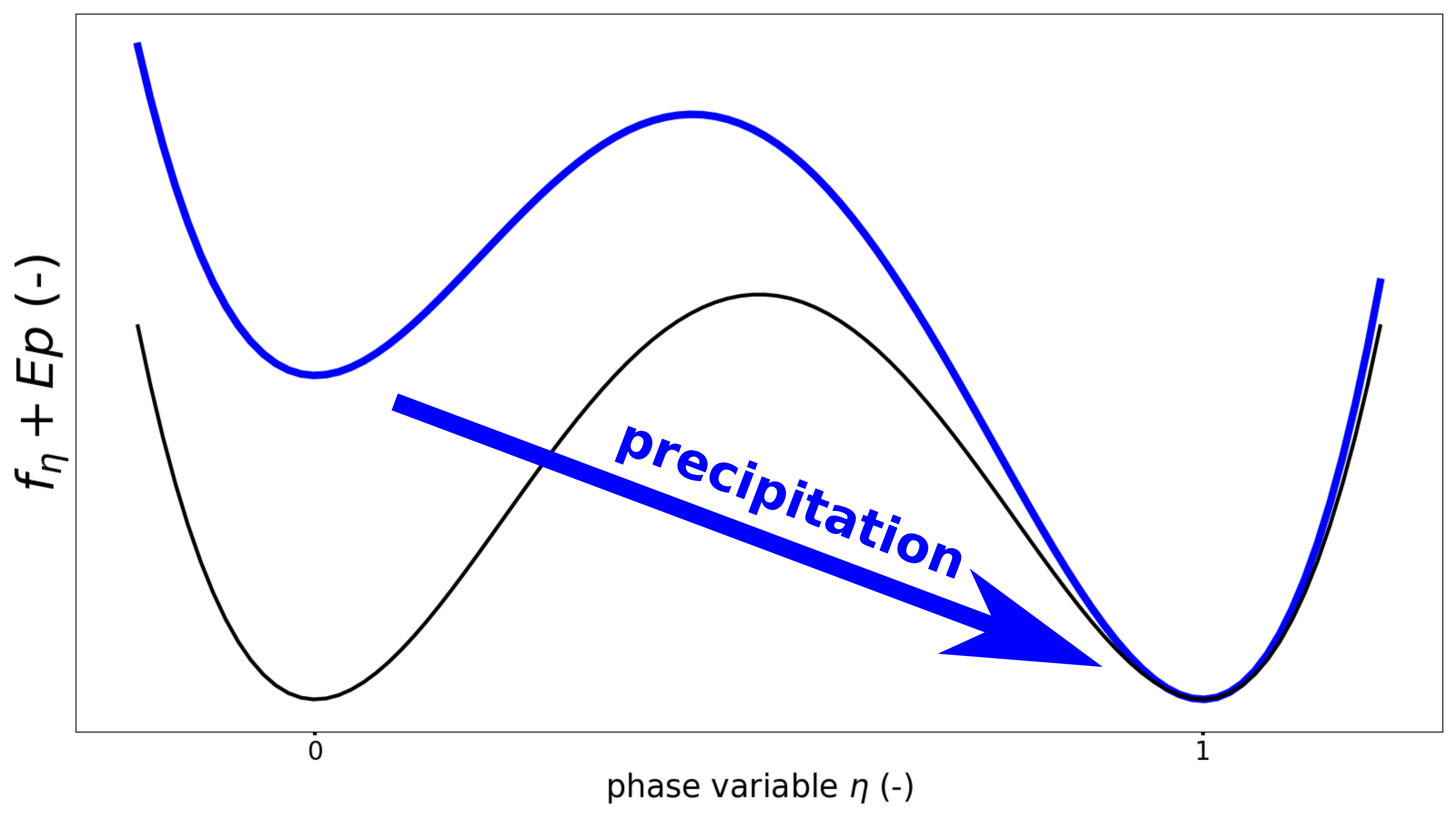}
    \caption{Example of tilted free energy $f_{CSH}+Ep(c, {CSH})$ (in blue), the precipitation is favored ($c>c_{eq,\,CSH}$).}
    \label{tilted f_phi Figure}
\end{figure}

\begin{equation}
    Ep(c, {CSH}) = \chi_{c\,, {CSH}}\times (c-c_{eq,\,CSH}) \times h_{CSH}({CSH})  
    \label{f_phi Ep Equation}
\end{equation}
Where, $h_{CSH}({CSH})=1-(3\,{CSH}^2-2\,{CSH}^3)$ is an interpolation function on the phase variable ${CSH}$ \cite{Takaki2014}, $\chi_{c,\, {CSH}}$ is the tilting coefficient related to the precipitation kinetics. This coefficient can be calibrated with experimental data.
Similarly, the choice of the interpolation function $h_{CSH}$ has an insignificant impact, while it verifies $h_{CSH}(0) = 1$, $h_{CSH}(1) = 0$, and $\frac{\partial h_{CSH}}{\partial CSH}(0) = \frac{\partial h_{CSH}}{\partial CSH}(1) = 0$.
It is important to notice the tilt occurs only if $c\ne c_{eq,\,CSH}$ and precipitation occurs in the case $c > c_{eq,\,CSH}$. The value $c_{eq,\,CSH}=0$ is considered for the equilibrium for the C-S-H precipitation. Figure \ref{tilted f_phi Figure} illustrates an example of tilting. 
It is important to point out that the term $\chi_{c\,, {CSH}}\times (c-c_{eq,\,CSH})$ determines the amplitude of the tilting, and so the
kinetics of the precipitation. The larger the tilting is, the faster the precipitation is.

Even if the model can predict the distribution of the volume fractions of the different phases in the domain, it is worth noting that the chemical shrinkage is not captured herein.
This phenomenon is defined as a reduction of the source+paste+water volume and an increase of the air volume (in proportion to conserve the volume of the sample) \cite{Powers1946}.
For simplicity reasons, the air phase is neglected and the domain is assumed to be always saturated (addition of water during the hydration).
Indeed, the effect of the shrinkage (and the required addition of water) remains small during the hydration phenomenon.

The PF framework with dissolution/precipitation is then ready to be employed to predict the microstructure evolution.


\section{Numerical microstructure evolution}
\label{Microstructure Evolution Cement}

Similarly to \cite{Nguyen2024}, two configurations are considered, employing a $100\,\mu m$ domain size: $w/c=0.3$ and $w/c=0.5$. Indeed, a saturation limit $w/c=0.42$ has been assumed to formulate the problem in Equation \ref{Mass Conservation Relation}, and the under- and over-saturation configurations are explored. Even if the framework is easily extended to 3D, only 2D slices have been considered in this paper. Indeed, the PF description can be computationally expensive. 

The different parameters are available in Table \ref{Parameter Simulation Microstructure}. 
Considering the lack of information on the value of the tilting coefficients $\chi_{c\,, {CSH}}/\chi_{c\,, {C3S}}$, they have been assumed equal.
Moreover, the competition between precipitation/dissolution $\chi_{c\,, {CSH}}/\chi_{c\,, {C3S}}$ and diffusion $\kappa_{c0}$ is primordial. Indeed, the slowest physics determines the global hydration kinetics.
It appears that the diffusion in the C-S-H gel dictates the rate of the hydration reactions \cite{Bullard2011, Scrivener2015}. 
This diffusivity value $\kappa_{c0}$ is uncertain and has been determined by a trial-and-error method to correlate data from \cite{Nguyen2024}.
At the contrary, the values of the tilting coefficients $\chi_{c\,, {CSH}}/\chi_{c\,, {C3S}}$ have an insignificant impact on the results as the precipitation/dissolution physics is faster than the diffusion.

The mesh size ($0.2 \, \mu m$) has been determined to limit the computation cost induced by the PF description, while ensuring the quality of the result. Such as comparison, a mesh size of $1\,\mu m$ is employed in the CA formulation \cite{Nguyen2024}.  
To investigate if the modelization is representative, at least ten simulations for each $w/c$ configuration have been conducted.

\begin{table}[ht]
    \centering
    \begin{tabular}{|l|r|}
        \hline
        Domain size & $100\, \mu m$ \\
        Mesh size $\Delta$ & $0.2\, \mu m$\\
        \hline
        Tilting coefficients $\chi_{c,C3S}/\chi_{c,CSH}$ & $0.1\,W_\eta$\\
        \hline
        Diffusivity $\kappa_{c0}$& $1\, \mu m^2$\\
        \hline
    \end{tabular}
    \caption{Parameters used during the microstructure evolution simulations.}
    \label{Parameter Simulation Microstructure}
\end{table}

\subsection{Initial conditions}
\label{IC Section}

Spheres are generated in a periodic domain, considering a given particle size distribution function \cite{Nguyen2021, NIST2005} and a random position avoiding overlap. This particle size distribution is modified, see \ref{PSD source}, as the nature of the PF theory limits the size of the particles ($R_{min}\geq 12\Delta$, where $\Delta$ is the mesh size).
Even if it is unclear in \cite{Nguyen2024}, such a modification seems also to be employed during the generation of the initial conditions for the CA approach. 
Spheres are generated until the domain reaches the targeted ratio $w/c$. 
It is worth noting that sphere shapes have been used for simplification reasons. However, the PF formulation can capture any heterogeneous shape, such as the irregular shapes obtained during the microstructure evolution depicted in Figure \ref{Microstructure Evolution Maps}. 

Once the position of the source of cement has been determined, it is possible to determine the map of the phase variable $C3S$. A signed distance to the interface water-source is computed and a cosine profile is used \cite{Takaki2014}, defined in Equation \ref{Cosine Profile}.

\begin{align}
        &=1 \text{ if } d \leq - \delta/2 \text{ (inside the grain)}\nonumber\\
    C3S&=0.5\left(1+\text{ cos}\left(\pi\frac{d+\delta/2}{\delta}\right)\right)\text{ if } |d|<\delta/2\label{Cosine Profile} \text{ (in the interface of the grain)}\\
        &=0 \text{ if } d \geq \delta/2 \text{ (outside the grain)}\nonumber
\end{align}
Where, $d$ is the signed ($-$ if inside the grain, $+$ if outside the grain) distance of the node to the interface and $\delta$ is the interface thickness.

Concerning the map of the phase variable $CSH$, one would like to set $CSH=0$ (no gel) everywhere. Then, the gel needs to nucleate and propagate with the consumption of the solute $c$. Nevertheless, PF formulation has some difficulties in capturing the nucleation phenomenon, in particular if it is localized. 
The physical mechanisms governing the initial formation of C-S-H are known to be complex, involving processes like frustrated nucleation and growth.
To ensure the localization of the nucleation around the source particle \citep{Taylor1997, Zhang2020}, a given number of nucleation seeds are therefore generated in the simulation. The same strategy was used in Molecular Dynamics precipitation simulations for the nucleation and growth of C-S-H \cite{Ioannidou2014, Goyal2020}. Their positions are randomly selected in the source neighborhood. Then, a cosine profile, see Equation \ref{Cosine Profile}, is applied with a radius of several mesh sizes. Furthermore, some noise is introduced into Equation \ref{Eq on phi} to obtain heterogeneous nucleation/growing due to imperfections or temperature fluctuations \citep{Petersen2018, Castro2003}.

Finally, the solute concentration $c$ is set to $c_{eq,\,CSH}=0$ in the domain, the value of the gel-water equilibrium. Without source grain, the domain is at the equilibrium. After the introduction of the source in the domain, dissolution occurs as the source-water equilibrium is not verified ($c=0$ versus $c_{eq,\,C3S}=1$).
The consequence of this dissolution is the increase of the solute concentration $c$ in the domain. With $c>c_{eq,\,CSH}$, the gel-water equilibrium becomes not verified, and precipitation occurs. As depicted in Figure \ref{Cementation Scheme Figure}, these processes verify the hydration phenomenon.


\subsection{Results}

\begin{figure}[ht]
    \centering
    \includegraphics[width=.7\linewidth]{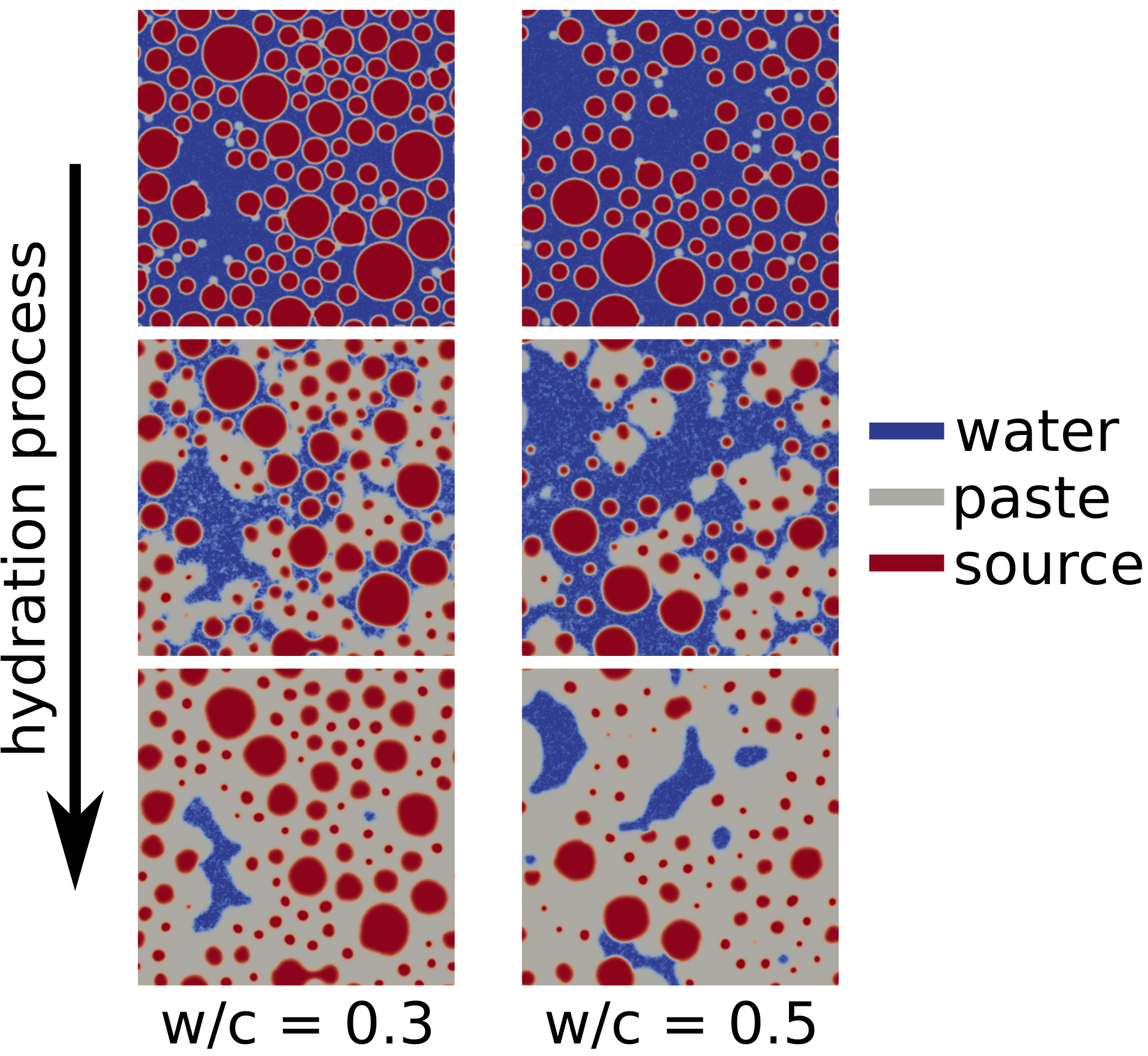}
    \caption{Evolution of the microstructure during the hydration process.}
    \label{Microstructure Evolution Maps}
\end{figure}

Figure \ref{Microstructure Evolution Maps} depicts examples of microstructure evolution, and different movies are available in supplementary documents. As expected, the source phase dissolves to produce the gel phase. Occasionally, a source grain may become entrapped within the cementitious matrix, subsequently acting as an inclusion within the hardened paste.
Indeed, the solute generated from the source dissolution is trapped in the paste. Remember that the diffusivity of the solute is much smaller in the paste than in the pore \cite{Bullard2011,Scrivener2015,Sun2022}, see Equation \ref{Solute Diffusivity}. As the solute is trapped, the solute concentration $c$ reaches the equilibrium value of the dissolution reaction $c_{eq,C3S}$, and the dissolution stops locally.
Furthermore, it appears the paste is not the only phase to precipitate (see at the bottom for $w/c=0.3$ in Figure \ref{Microstructure Evolution Maps}). Indeed, the source phase precipitates also. 
The PF theory can induce curvature-driven shape evolution to reduce surface tension \cite{Moelans2008, Allen1979}. The origin of this unphysical phase evolution is the gradient coefficient $\kappa_\eta$ from Equations \ref{Eq on psi} and \ref{Eq on phi}. The solution to this problem appears to be the use of new distinct phase variables for each grain, increasing the computational cost \cite{SacMorane:PFDEM}.
The volume fraction of this reprecipitated source was found to be less than 0.3\% in all simulations (mean value of 0.1\%) and thus has a minimal effect on the overall phase fractions and computed mechanical properties.

It is worth noting that microstructures from Figure \ref{Microstructure Evolution Maps} have not reached the final configuration. Indeed, some chemical reactions occur until the total consumption of the source particle (for the undersaturated configuration, $w/c=0.5$) or until the saturation state is reached (for the oversaturated configuration, $w/c=0.3$).
However, the hydration kinetics are strongly slowed down by the trapping of the solute in the paste. The final configurations depicted herein are expected to be a sufficient approximation of the steady state. 

Compared to microstructures obtained with a \emph{CEMHYD3D} model \cite{Nguyen2024} or observed after a real hydration experiment \cite{Thomas2009}, the phase boundaries are smoother and the integrity of the phases is conserved in the PF framework. Indeed, the PF framework has been observed to facilitate the smoothing of interfaces \cite{Kelm2022}, which may lead to the inaccurate capture of the fingering growth pattern \cite{Thomas2009}. A finer mesh would help to capture heterogeneous shapes, increasing the computational cost.

This chemical process can be described by the degree of hydration of the source material $h$, defined in Equation \ref{Equation Hydration}. This index varies between $0$ for intact source particles initially to $1$ when all the source phases are dissolved.

\begin{equation}
    h=\frac{C3S_i-C3S_t}{C3S_i}
    \label{Equation Hydration}
\end{equation}
Where, $C3S_i$ and $C3S_t$ are the source variable's initial and current mean values.

\begin{figure}[ht]
    \centering
    \includegraphics[width=0.6\linewidth]{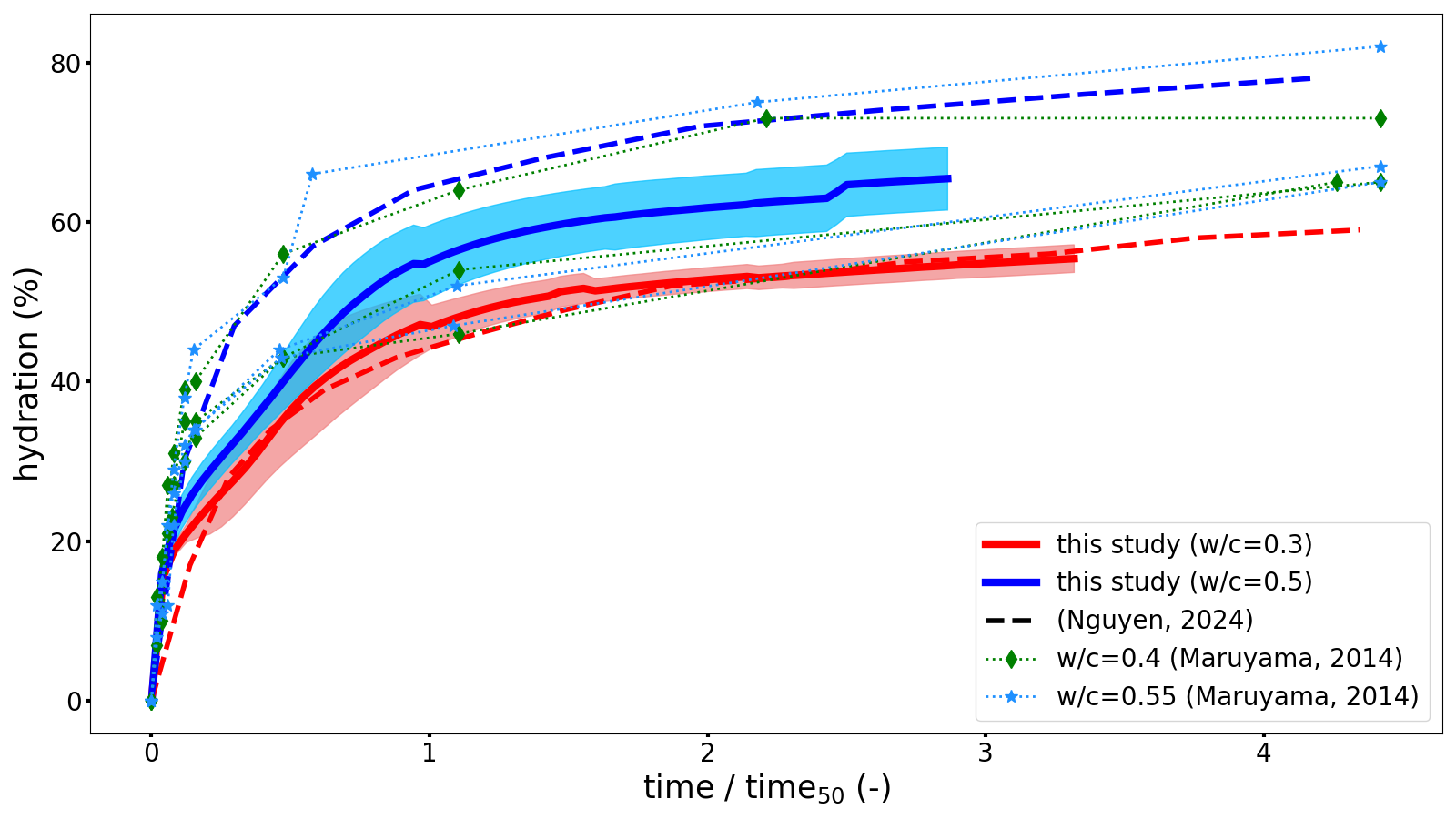}
    \caption{Time evolution of the degree of hydration in the distinct cases (configurations are repeated ten times). $time_{50}$ is the normalization factor defined as the mean time required to reach $h=50\%$.}
    \label{Time Hydration}
\end{figure}

Figure \ref{Time Hydration} shows the evolution of the hydration in the two cases ($w/c=0.3$ and $0.5$). The results are compared to the hydration curves obtained with the CA description in \cite{Nguyen2024} and to experimental results obtained with $w/c=0.4$ and $=0.55$ \cite{Maruyama2014}.
The time has been normalized to allow the comparison between the results.
Considering the repetitions of the configurations, Figure \ref{Time Hydration} depicts the average value in plain lines and an envelope in the color areas. This envelope considers the extrema values $\mu-\sigma$ and $\mu+\sigma$, where $\mu$ is the mean value and $\sigma$ is the variance at a given x-coordinate. This representation is preserved in the following of this work.

Hydration appears to be divided into two parts: a fast dissolution period and a slow dissolution period \cite{Bullard2011,Scrivener2015,Sun2022}.
In the first part of the curve, dissolution occurs. However, the diffusivity of the solute aims to decrease with the precipitation of the C-S-H, and the solute concentration locally reaches the water-source equilibrium value. This phenomenon slows down the dissolution of the source, explaining the second part of the curve.
As expected, the hydration kinetics are faster for the configuration $w/c=0.5$ than for $w/c=0.3$. Indeed, $w/c=0.5$ is an under-saturation configuration, and the dissolution of the source particles is favored.
The results from the PF simulations appear to correlate well with the CA prediction and with the experimental observations (even if the $w/c$ are not similar).
About the representativeness of the simulations, the results for the $w/c=0.3$ configuration are strongly similar. In the same vein, the results for the $w/c=0.5$ configuration appear reproducible, even if the variance is a bit larger. 
This aspect can be diminished, considering a larger domain size. In the following, the $100\,\mu m$ domain is conserved for comparison purposes with \cite{Nguyen2024}.

\begin{figure}[ht]
    \centering
    a) \includegraphics[width=0.6\linewidth]{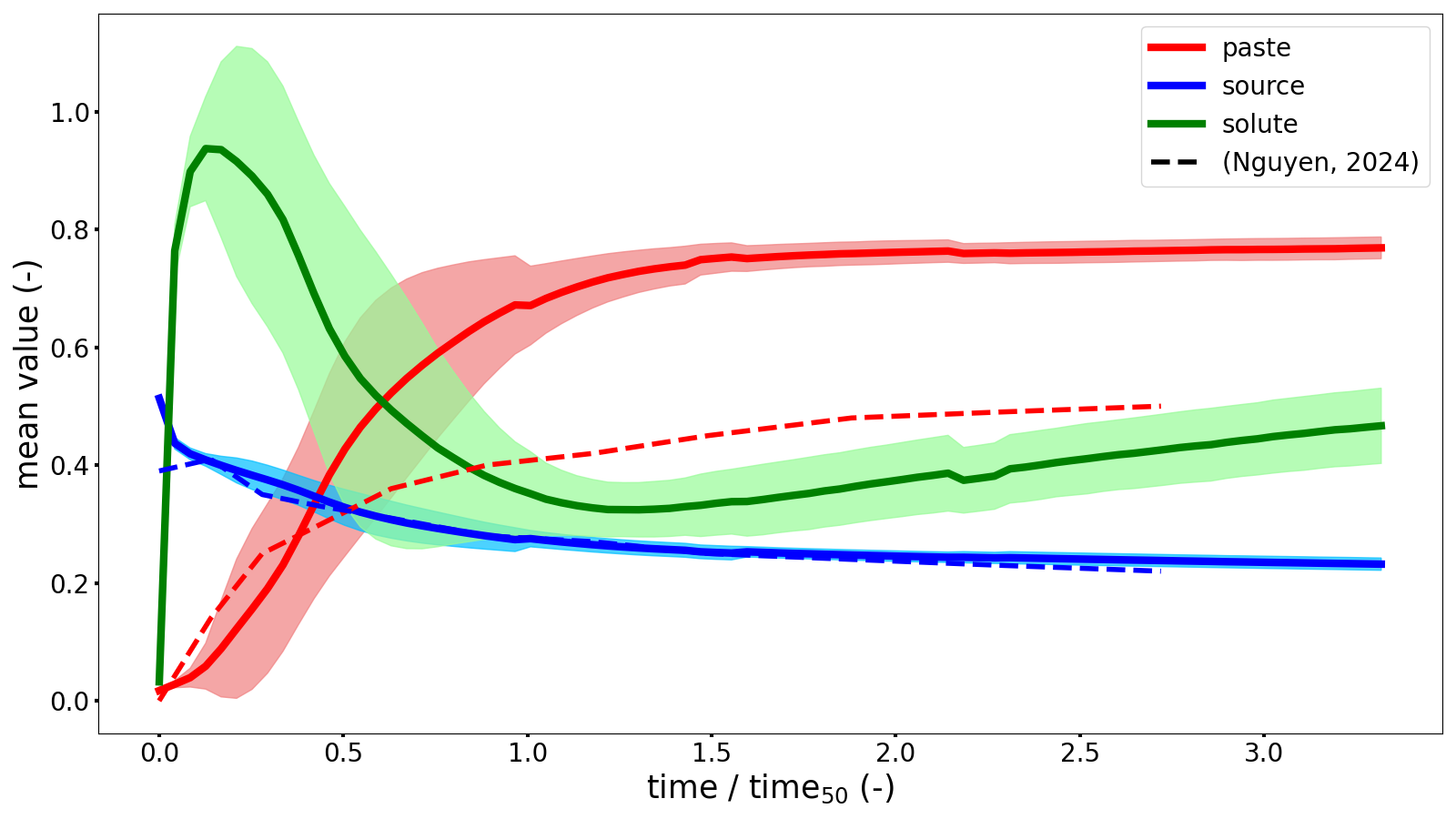}\\
    b) \includegraphics[width=0.6\linewidth]{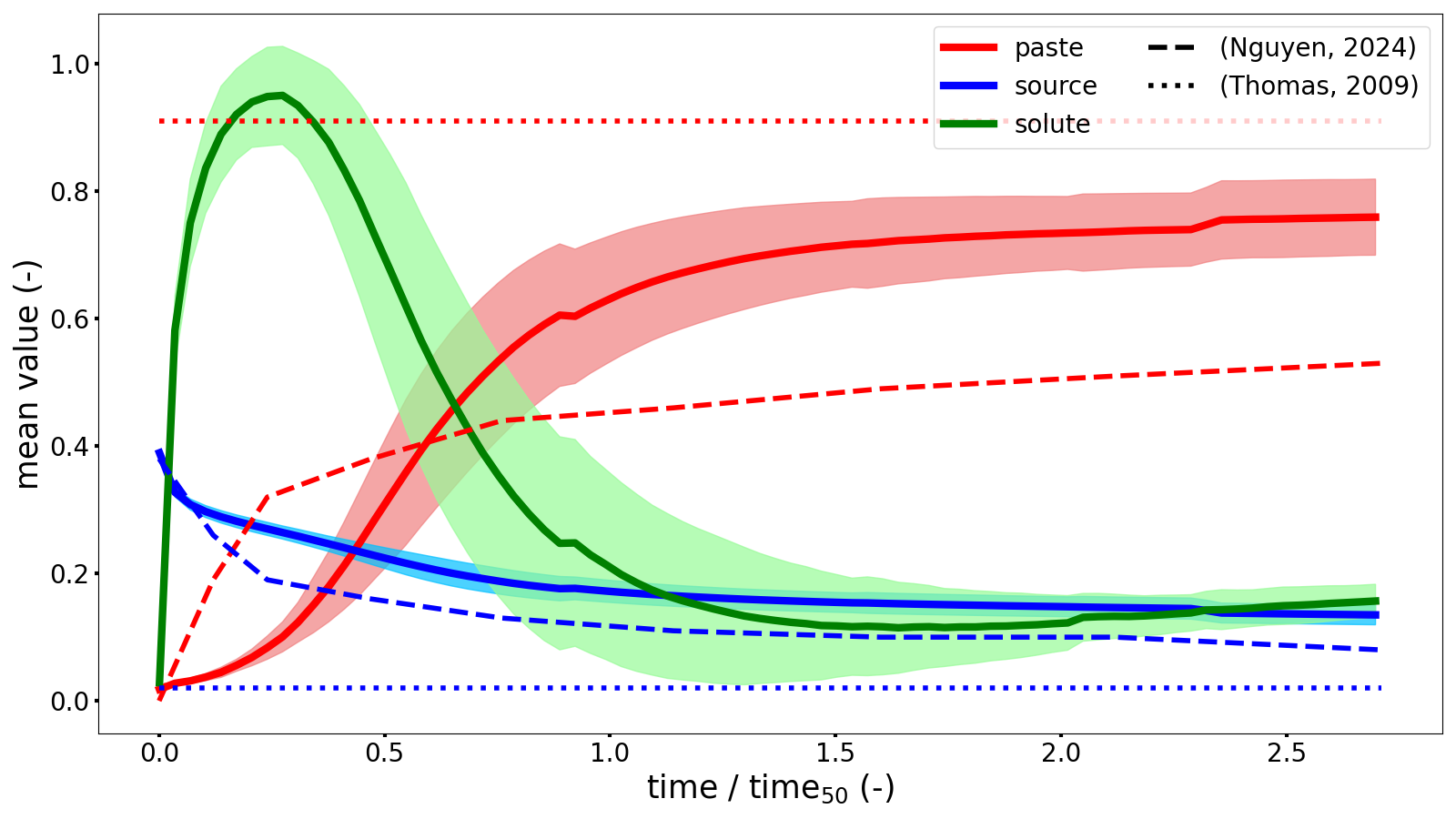}
    \caption{Time evolution of the different volume fractions for the species (source, paste, solute) in the case $w/c=$ a) $0.3$ and b) $0.5$ (configurations are repeated ten times).}
    \label{Time Mean Value Species}
\end{figure}

Then, the microstructure evolution can be described through the evolution of the volume fractions of the different species, see Figure \ref{Time Mean Value Species}. The hydration phenomenon is well divided into the dissolution of the source to generate some solute and then the consumption of the solute to precipitate the paste, with an intermediary supersaturation of the pore space with respect to the gel.

If the mean value of the source quantity appears similar to the one obtained by \cite{Nguyen2024} ($C_3S,\,C_2S,\,C_3A,\,C_4AF$ species are considered), it is not the case for the mean value of the paste quantity ($CSH,\,CH$ species are considered), which is larger in the PF simulations. 
Even if supplementary phases (available in \cite{Nguyen2024}) were included such as the gypsum, the anhydrite, the ettringite, the $AFm$ or $FH_3$, these phases represent only $\sim 19\,\%$ of the volume for $w/c=0.3$ and $\sim 16\,\%$ of the volume for $w/c=0.5$ \cite{Haecker2005}. 
It is not sufficient to fill the gap between the PF and the CA descriptions.
In order to determine the most accurate formulation, the final concentrations of the source and the paste obtained from the hydrated microstructure available in \cite{Thomas2009} ($w/c=0.5$) are depicted for comparison.
The results obtained with the PF description appear more accurate.
It seems that \emph{CEMHYD3D} simulations overestimate the pore space, in particular for the case $w/c=0.5$.
Furthermore, it is not possible to estimate the mean value of the solute concentration in the data described in \cite{Nguyen2024}. Indeed, the \emph{CEMHYD3D} description does not consider this specific phase. 
In the CA approach, the source phase is transformed directly into a paste phase if the chemical conditions are set (reaction between the given phases).
Compared to the data from \cite{Nguyen2024}, a delay can be appreciated in the PF formulation concerning the precipitation of the paste, due to the required diffusion of the solute. 
Similarly to the hydration curves, the PF simulations appear reproducible in terms of phase distribution, in particular for $w/c=0.3$. Beyond these bulk volume fractions, a deeper understanding of the microstructure's development can be gained by analyzing the morphology and spatial organization of the phases. 

\begin{figure}[ht]
    \centering
    \includegraphics[width=0.6\linewidth]{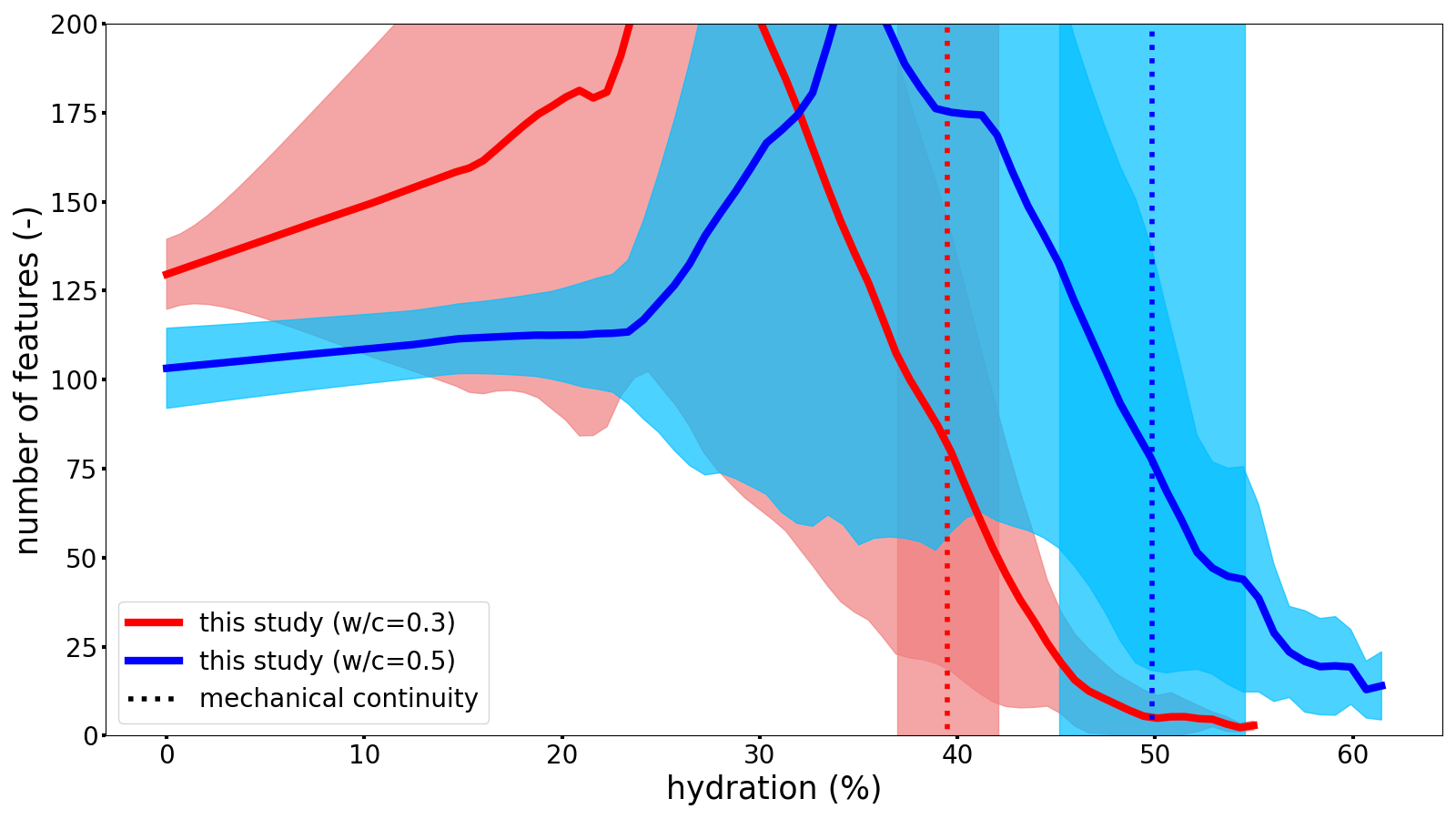}
    \caption{Evolution of the number of features for the phase paste+source.}
    \label{Hydration Segmentation}
\end{figure}

To do so, the microstructure is segmented into three phases: paste, source, and paste+source. The different nodes of the mesh are considered in the paste phase if $CSH>0.5$. Similarly, the nodes are considered in the source phase if $C3S>0.5$. The paste+source phase is a combination of the two previous phases.
Then, the connectivity of the phases is quantified by labeling the different phases of the segmented images with the Python function \emph{scipy.ndimage.label}. This operation distinguishes the individual features from a given phase.
A feature is defined as a phase island surrounded by the pore/water.
For instance, the number of features for the paste+source phase is depicted in Figure \ref{Hydration Segmentation}. It appears this number increases and then decreases as the source dissolves. 
The initial increase is due to the nucleation of the cement paste, while the reduction occurs as the source grains dissolve and the distinct paste seeds interconnect.
It is worth noting the wide envelopes of this descriptor. 
Indeed, this specific morphometer remains at the microscale, counting the individual features, while the others consider a homogenization at the sample scale, computing an average value.
It remains sensitive to the microstructure evolution, especially considering a domain size of $100\,\mu m$.

\begin{figure}[ht]
    \centering
    \includegraphics[width=0.9\linewidth]{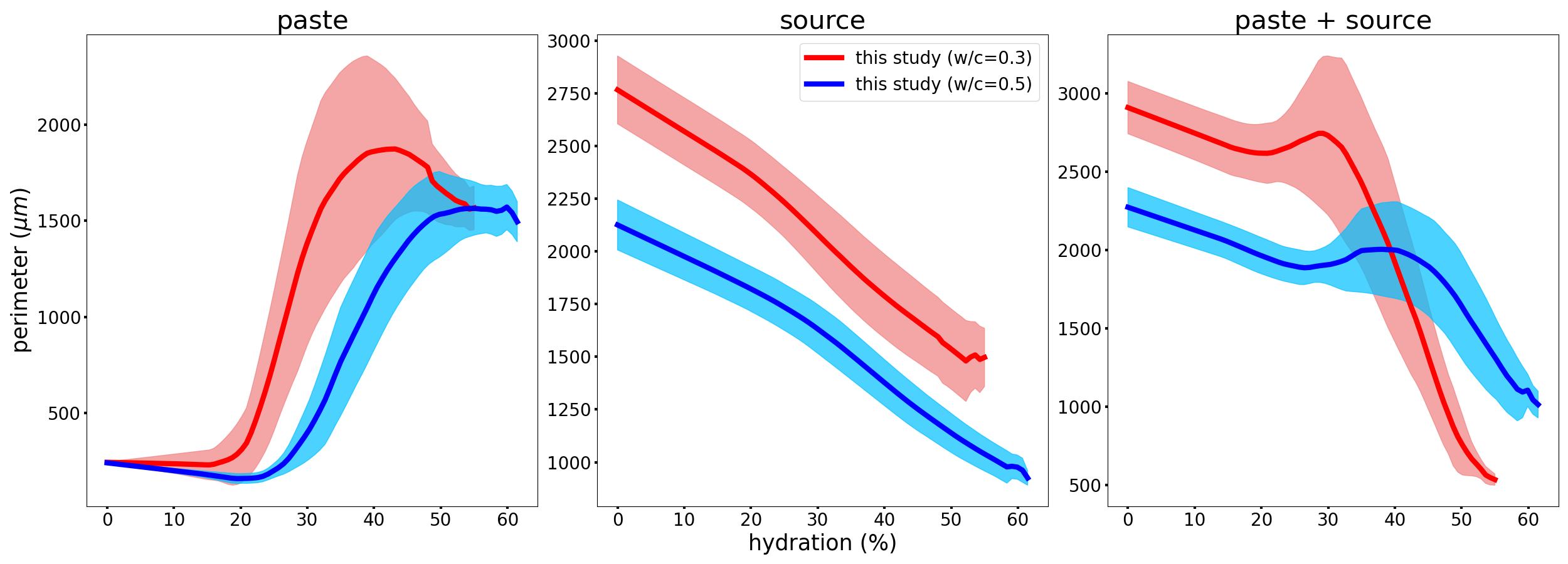}\\
    \caption{Evolution of the perimeter of the different phases (paste, source, paste+source) during the hydration (configurations are repeated ten times).}
    \label{Hydration Perimeter}
\end{figure}

The microstructure can be more deeply described by investigating the evolution of the perimeter of each phase (paste, source, paste+source), see Figure \ref{Hydration Perimeter}. 
This perimeter index is obtained by the function \emph{measure.perimeter} from the Python module \emph{skimage}. 

The perimeter of the source phase appears to decrease with the hydration (dissolution). On the contrary, the paste perimeter increases with the hydration (precipitation). Similarly to the phase distribution curves, the delay between the dissolution and the precipitation can be pointed out. Indeed, the perimeter of the paste does not evolve initially, while the perimeter of the source decreases, even at the initial times.
Furthermore, a decrease in the paste perimeter during the hydration can be noticed at the end of the simulation. Indeed, as illustrated in Figures \ref{Microstructure Evolution Maps}, the different paste seeds can interconnect, reducing the phase perimeter. 
This phenomenon also explains the slope variation of the paste+source perimeter curve. The decrease in the perimeter for the combined phase accelerates with the interconnection of the seeds. 
The nucleation and the coalescence of these aggregates are also depicted in Figure \ref{Hydration Segmentation}, where the number of the distinct islands (paste+source phase surrounded by pore) shows an increase followed by a decrease during the hydration process.
Similarly, Figure \ref{Time Mean Value Species} and \cite{Ioannidou2014} emphasize the fact that the C-S-H volume fraction evolves with a sigmoidal pattern.
While a direct overlay of the plots would be inappropriate as they measure different quantities (morphology vs. bulk volume), the non-monotonic hump shape of the perimeter curve is the direct morphological signature of the sigmoidal growth, revealing the nucleation/growing of separate islands followed by the coalescence of these aggregates.
In the same vein as previously, the microstructure obtained with the PF formulation appears reproducible, regarding the size of the envelope.

\begin{figure}[ht]
    \centering
    a) \includegraphics[width=0.8\linewidth]{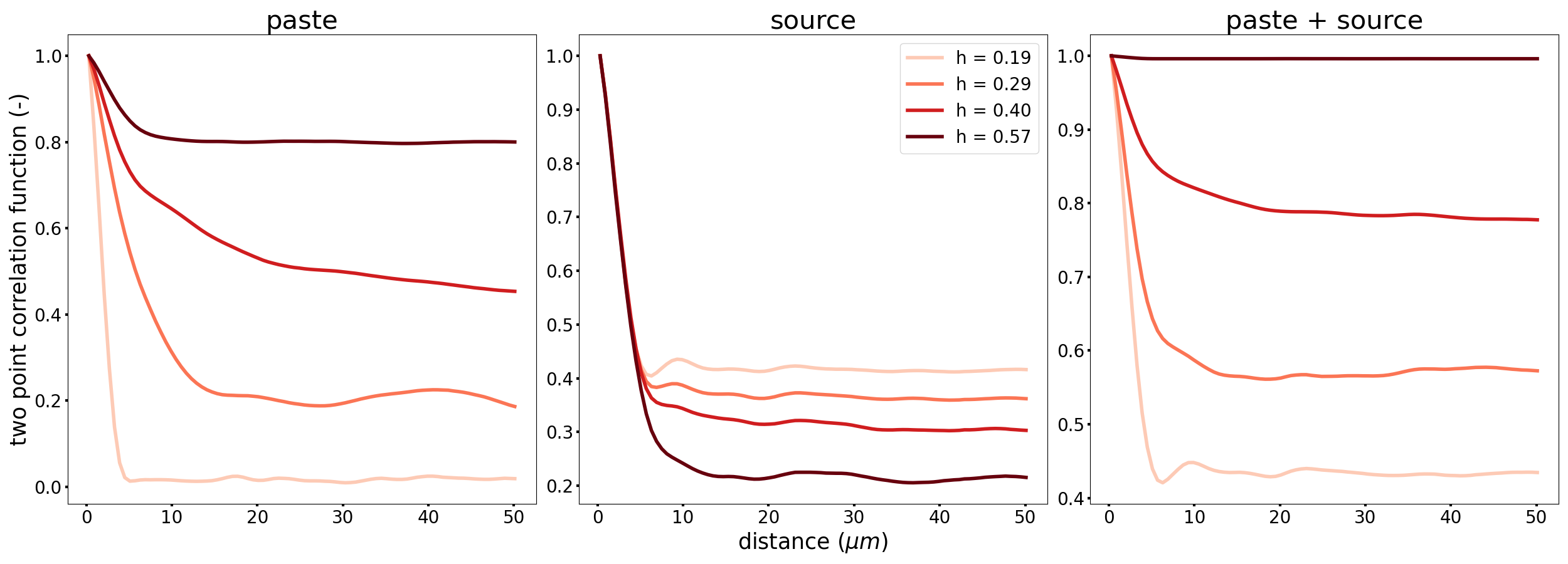}\\
    b) \includegraphics[width=0.8\linewidth]{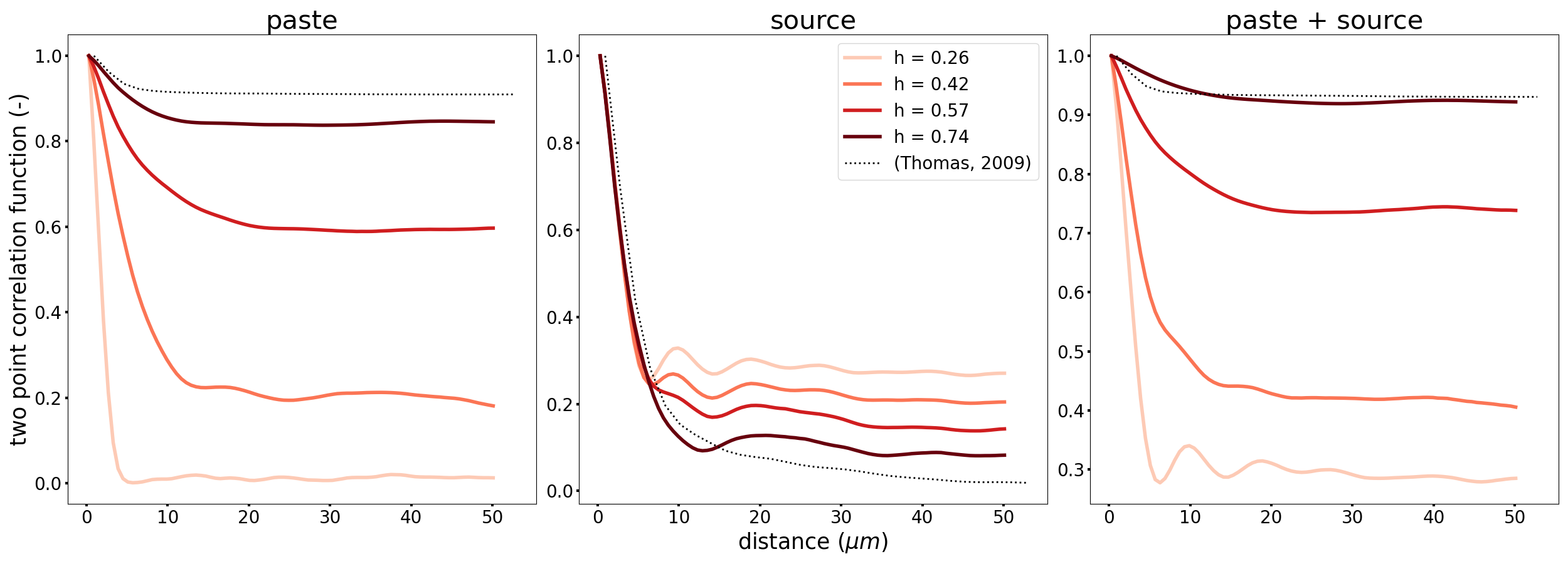}
    \caption{Evolution of the correlation function for the different phases (paste, source, paste+source) in the case $w/c=$ a) $0.3$ and b) $0.5$.}
    \label{Hydration Correlation}
\end{figure}

Subsequently, these segmented images can be used to compute the normalized two-point correlation function,  \cite{Bentz2006} and \ref{Definition Correlation} for a definition, see Figure \ref{Hydration Correlation} (the results of only one simulation for $w/c=0.3$ and $=0.5$ are depicted for clarity reasons; however, the repeatability is discussed in \ref{Correlation Repetition}). This tool adds details on the structure of the microstructure by giving the characteristic size of the phase. Indeed, this function takes a value between $1$ (highly correlated) and $0$ (not correlated) for different length dimensions. 
For example, it appears the paste correlation function drops at the initial condition ($h\rightarrow0$) for the distance equal to $5\,\mu m$ (initial size of the seeds). Then, the correlation function increases for the different sizes as the paste phase precipitates.
Similarly, the source correlation function decreases for all sizes with the hydration as the source dissolves.
For comparison with the case $w/c=0.5$, a real microstructure, from \cite{Thomas2009}, is characterized. The phase paste+source appears well described by the PF formulation. It is worth noting that the paste correlation is lightly underestimated, whereas the source correlation is lightly overestimated. 
These differences may be filled by considering a wider source size distribution or capturing the fingering growth pattern of the cement paste. 
Furthermore, the evolution of the two-point correlation function for the paste phase available in Figure \ref{Hydration Correlation} provides a quantitative signature of the developing microstructure. 
The progressive flattening of the curve at higher degrees of hydration indicates the establishment of long-range spatial correlations, which is the statistical marker of the percolation threshold.
This signifies the physical transition of the paste from a collection of isolated aggregates into a continuous, sample-spanning network. 
This mechanism is also visible through the increase-decrease evolution of the paste perimeter, see Figure \ref{Hydration Perimeter}, and through the decrease of the number of isolated material islands, see Figure \ref{Hydration Segmentation}.
This result is conceptually consistent with findings from more fundamental molecular dynamics studies of C-S-H formation, which also investigate the development of a cohesive, percolating gel network \cite{Ioannidou2016, Ioannidou2014}.
The model's ability to capture this critical topological transition therefore provides a validation of its physical realism.

Before the next Section, please note that other indices could have been applied to describe the microstructure \cite{Petersen2018, Guevel2022, Lindqwister2025, Torquato2002}, such as the chord-length density function, the pore-size function, the Euler number, or the specific surface. These indices would characterize the microstructure with complementary information.


\section{Mechanical characterization of the hydrated microstructure}

\subsection{Formulation of the mechanical homogenization}

\begin{figure}[ht]
    \centering
    \includegraphics[width=0.8\linewidth]{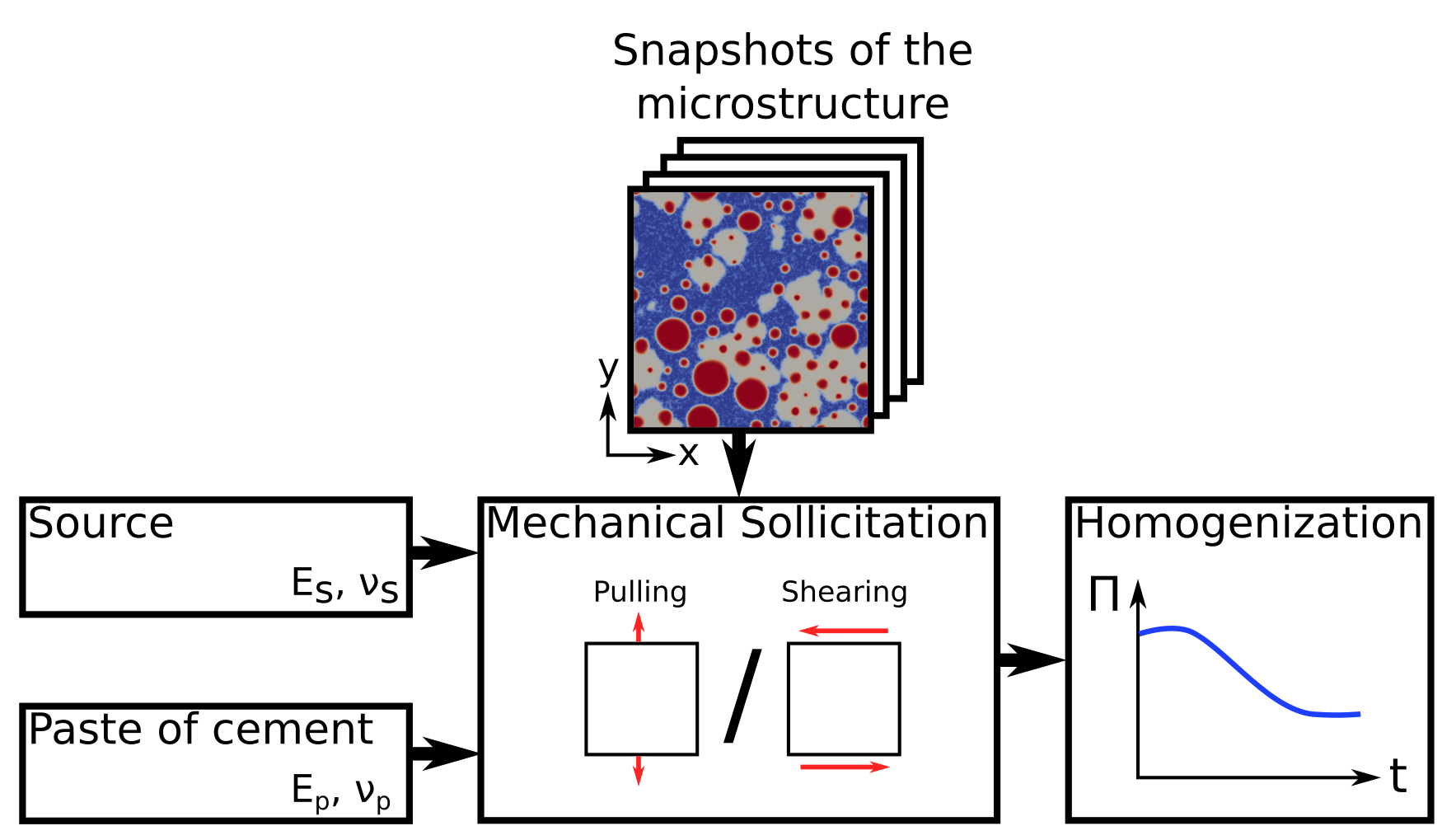}
    \caption{The microstructure is loaded to compute homogenized mechanical parameters.}
    \label{Homogenization Scheme}
\end{figure}

After estimating the microstructure evolution, mechanical loading (tensile and shear forces) is applied at the specimen's boundaries to determine the homogenized parameters, including Young’s modulus and Shear modulus, see Figure \ref{Homogenization Scheme}.
It is worth noting that the chemical aspect of the problem (microstructure evolution) is paused during this mechanical estimation of the composite material. 
The distinct microstructures obtained in Section \ref{Microstructure Evolution Cement} are employed as independent inputs in the following. 
Furthermore, the impact of the mechanics on the chemistry and microstructure evolution is neglected herein. However, this feedback influence is pivotal in some contexts, such as the underground storage (involving hydrating concrete wells under stress \cite{Dalton2022, Chavez2020, Fabbri2009}), and can be considered.
For instance, a stress-dependent microstructure evolution has already been proposed in \cite{Guevel2020}. In the same vein, \cite{SacMorane:PFDEM, SacMorane:PFDEMb} depicts the development of a new method to consider the impact of the stress and the microstructure reorganization on the chemistry.

A displacement (vertical in the case of pulling or lateral in the case of shearing) is applied on the top surface, whereas the bottom surface is assumed fixed. 
Periodic conditions are considered for the left and right surfaces to avoid any influence of the boundary conditions.
It is worth noting that it is important to verify the mechanical continuity of the sample before the interpolation of the mechanical characteristics.
To ensure force transmission between the top and bottom surfaces, the paste+source phase must be continuous, while the pore phase is not considered in the transmission of forces. Consequently, the initial microstructures formed during hydration are not used in the estimation of the sample’s mechanical parameters as this criterion is not verified. Figure \ref{Hydration Segmentation} depicts the mean hydration (and its variation) ensuring the mechanical continuity.

Three phases are identified and assigned to the mesh: the paste, the source, and the water in the pore, all having an isotropic elastic mechanical behavior described in Table \ref{Mechanical Parameter Phases} \cite{Nguyen2024,Ioannidou2016,Hu2014,Haecker2005}. 
It is worth noting that the mechanical behavior of the paste has been simplified in this initial estimation.
Indeed, it has been proved that the paste tends to behave such as a viscoelastic material, in particular at the early age \cite{Alizadeh2010,Hu2019,Yahia2023}. 
Considering the intrinsic viscosity of the paste, the model would be able to predict the creep behavior of the hydrated cement.
To ensure the water is not transmitting stress in the medium, the Young modulus assigned is small compared to the other phases. Hence, the stress will concentrate on the harder phases: the paste and the source.

\begin{table}[ht]
    \centering
    \begin{tabular}{|l|c|c|}
        \hline
        Phase & Young modulus (GPa) & Poisson ratio (-) \\
        \hline
        Paste & 25 & 0.24\\
        \hline
        Source & 120 & 0.3\\
        \hline
        Pore & 0.12 & 0.3\\
        \hline
    \end{tabular}
    \caption{Elastic properties considered for the different individual phases (from \cite{Nguyen2021}).}
    \label{Mechanical Parameter Phases}
\end{table}

\begin{figure}[h]
    \centering
    a) \includegraphics[width=0.45\linewidth]{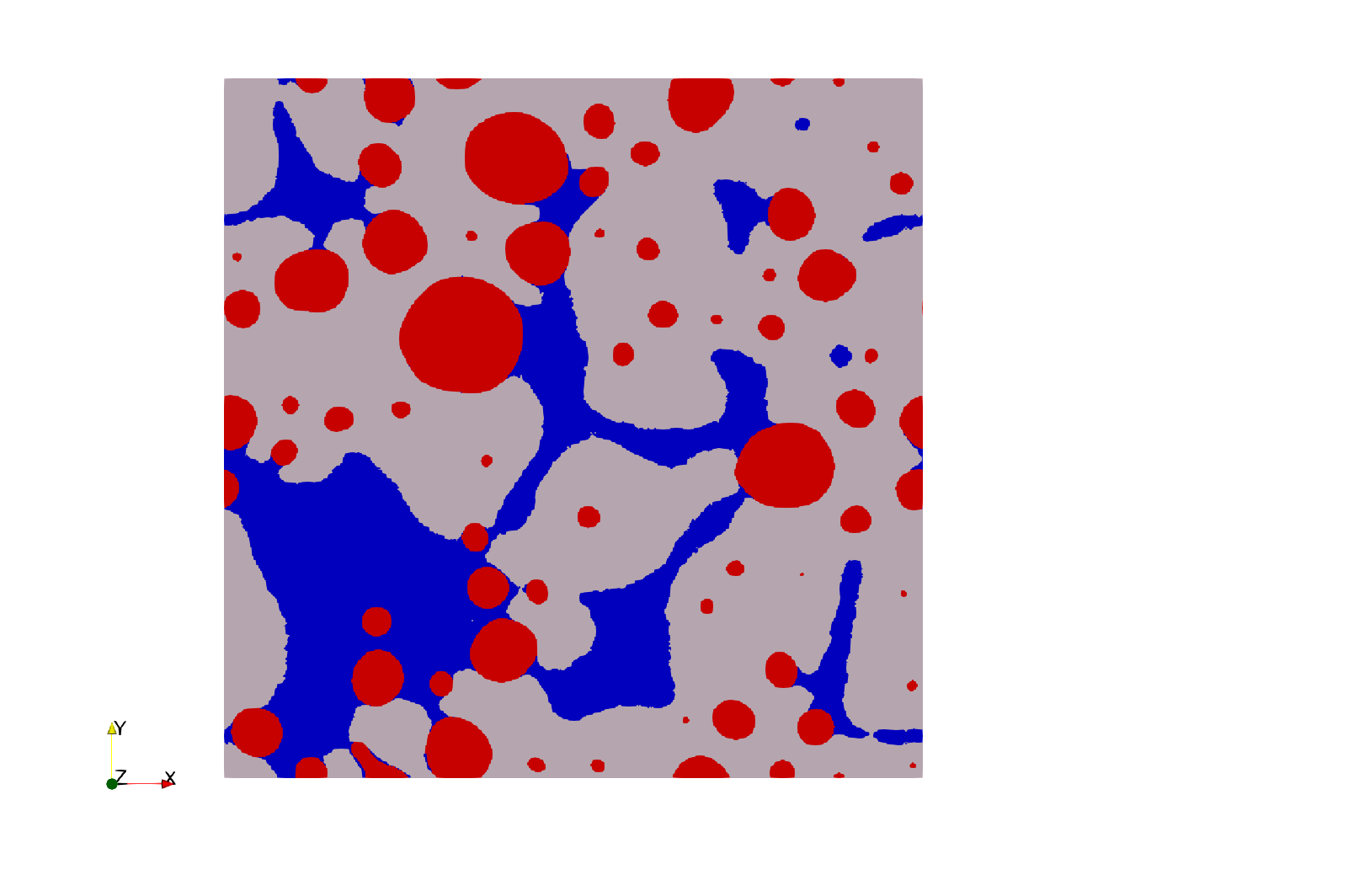}
    b) \includegraphics[width=0.45\linewidth]{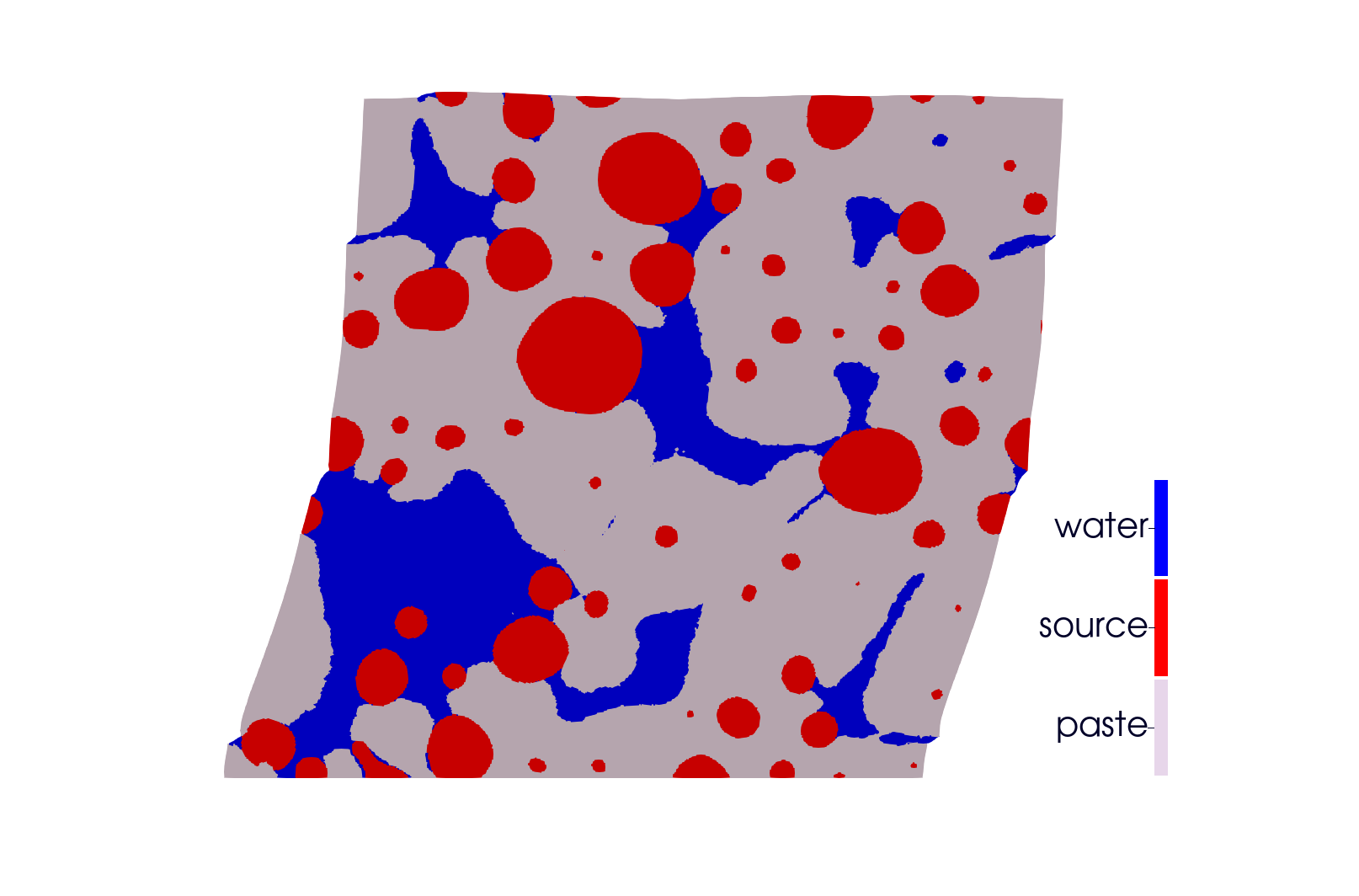}\\
    c) \includegraphics[width=0.6\linewidth]{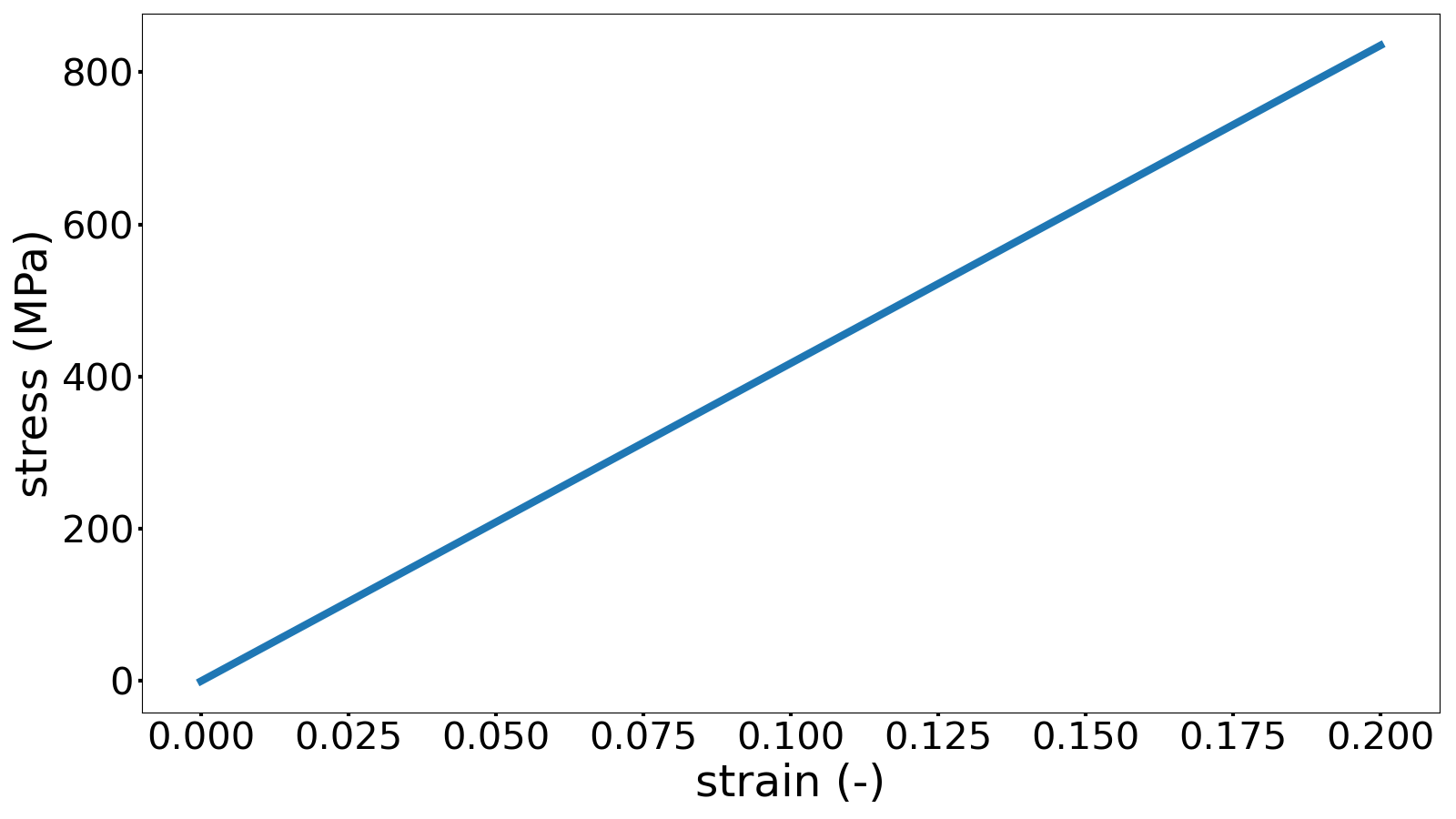}
    \caption{Example of the prediction of the shear modulus for a $w/c=0.5$ configuration: a) the composite material is b) loaded, and the parameter is estimated from c) the strain-stress curve.}
    \label{Mechanical Loading Example}
\end{figure}

The loading is applied until reaching $\epsilon_y=\Delta_y/height=0.2$ (pulling) or $\gamma=\Delta_x/height=0.2$ (shearing), where $\Delta_y$ is the displacement following the y-axis, $\Delta_x$ is the displacement following the x-axis, and $height$ is the dimension of the sample in the y-axis. 
The mechanical problem is also solved with the finite element solver MOOSE \cite{MOOSE}.
An example of microstructure loading is depicted in Figure \ref{Mechanical Loading Example}. 

Then, the mechanical parameters of the sample are computed by homogenization of the stresses and the strains in the domain.
The Young modulus $E_h$ is determined with the traction test by interpolating the relation $\sigma_{yy} = E_h \cdot \epsilon_y$, where $\sigma_{yy}$ is the mean stress in the source and paste phases and $\epsilon_y$ is the loading condition.
The Shear modulus $G_h$ is determined with the shearing test by interpolating the relation $\sigma_{xy} = G_h \cdot \gamma$, where $\sigma_{xy}$ is the mean stress in the source and paste phases and $\gamma$ is the loading condition.
The Poisson's ratio $\nu_h$ can be interpolated with a traction test without periodic conditions (not conducted herein) and the Bulk modulus $K_h$ can also be interpolated with an isotropic loading test (not conducted herein).
These parameters could also be interpolated from the interpolated Young and Shear moduli $\left(\nu_h= E_h/(2\cdot G_h)-1 \text{ and } K_h=(E_h\cdot G_h)/(9G_h-3E_h)\right)$.

The framework presented herein could be extended to determine numerically the yield envelope of the sample. 
To achieve this, each distinct phase should be characterized by a plasticity criterion, and the sample should be subjected to various mechanical loading paths until the overall mechanical behavior deviates from linearity (after significant plastic deformation occurs locally) \cite{Lesueur2021, Lesueur2022}.


\subsection{Results}

\begin{figure}[ht]
    \centering
    a) \includegraphics[width=0.45\linewidth]{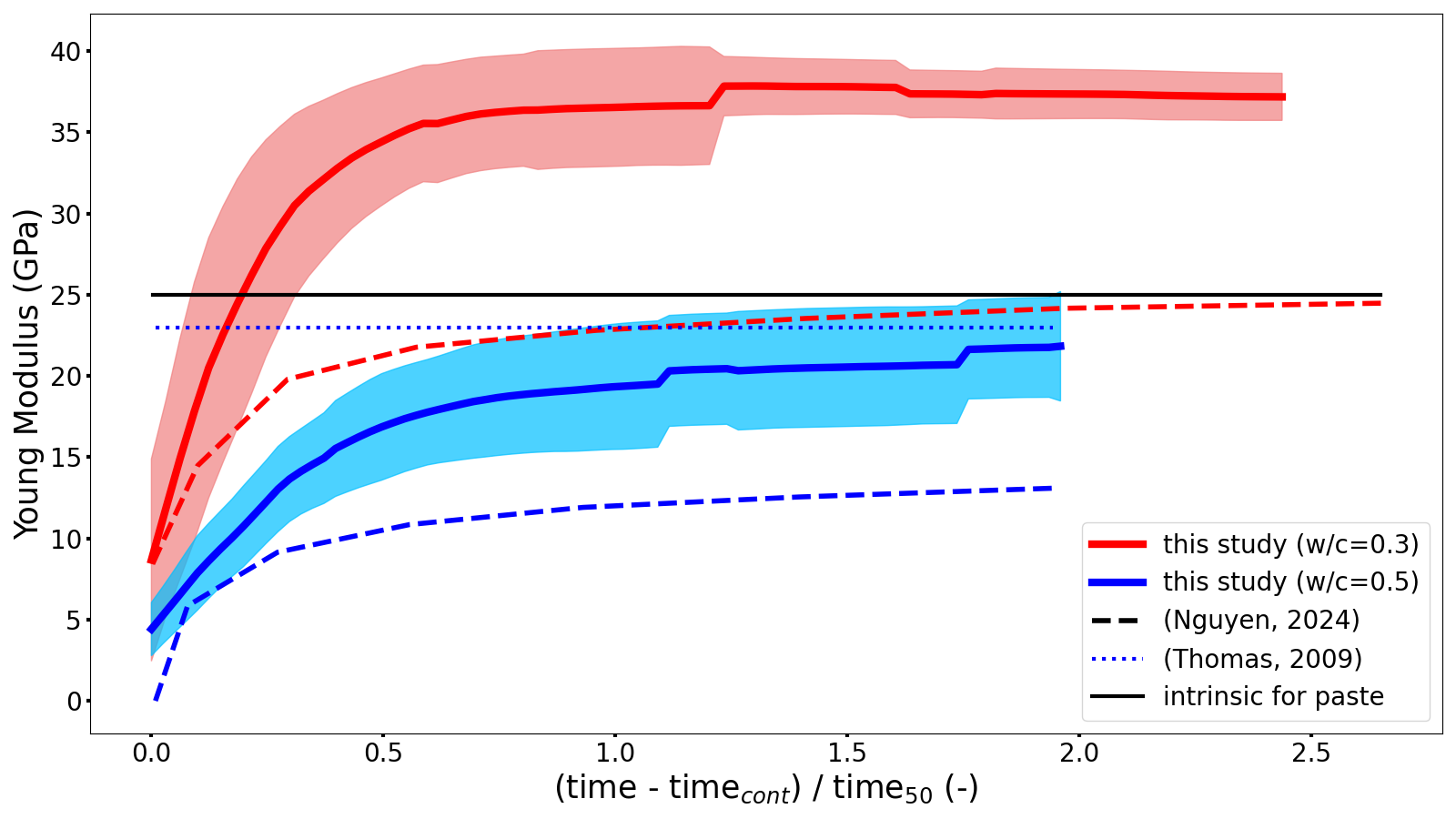} 
    b) \includegraphics[width=0.45\linewidth]{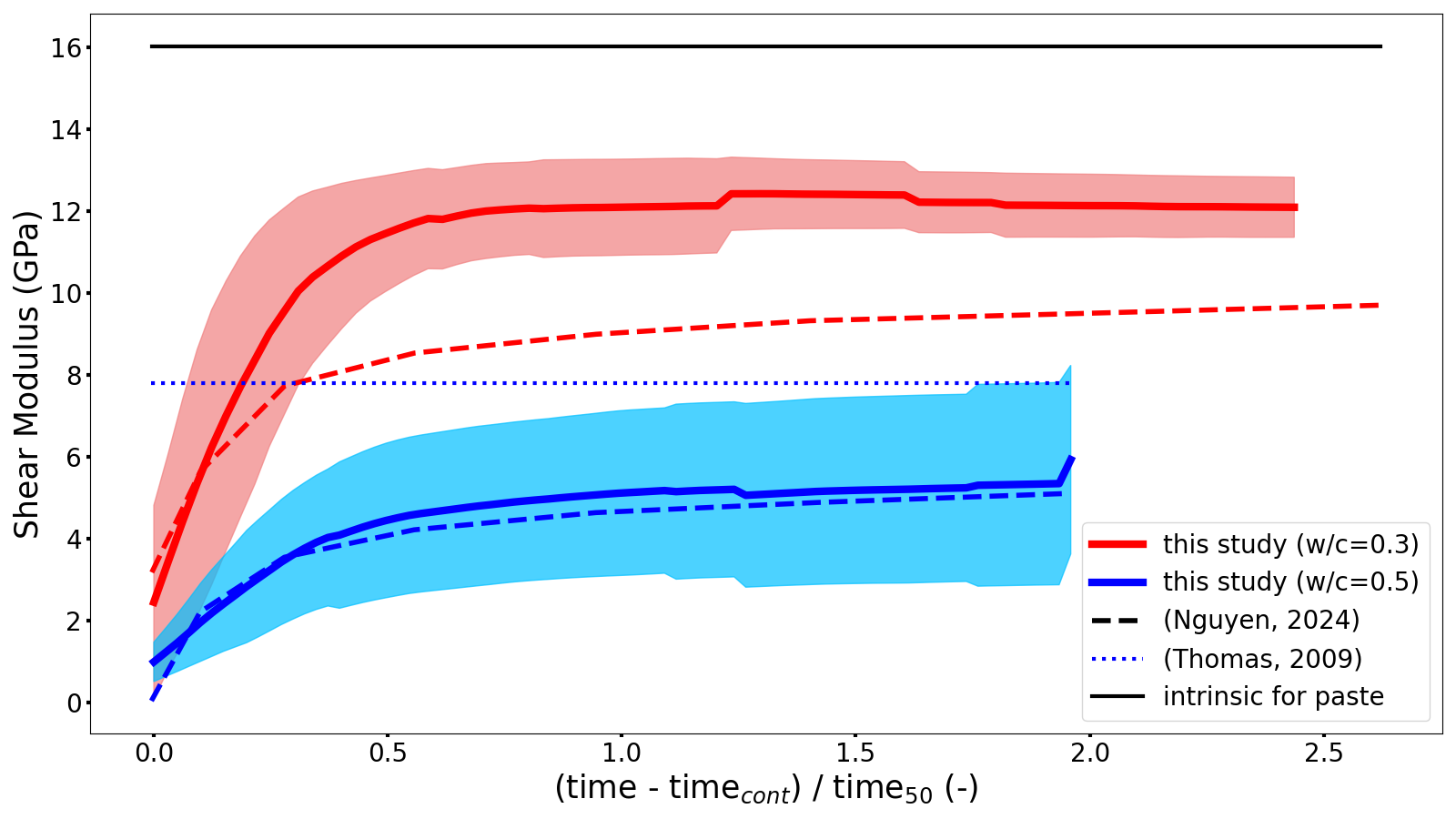}
    \caption{Time evolution of the homogenized a) Young modulus $E_h$ and b) Shear modulus $G_h$ in the cases $w/c=0.3$ and $0.5$.}
    \label{Mechanical Evolution}
\end{figure}

The evolution of the homogenized Young modulus $E_h$ and Shear modulus $G_h$ is available in Figure \ref{Mechanical Evolution}. The cases $w/c=0.3$ and $w/c=0.5$ are compared to the results from \cite{Nguyen2024} and the intrinsic value for the paste phase, dominant in the domain and representing the weaker phase (pore/water excluded). Moreover, a realistic microstructure obtained by Scanning Electron Microscope observation of a $w/c=0.5$ sample, from \cite{Thomas2009}, is used as an input of the homogenization method for comparison purposes.

It is worth noting that the hydration that ensures the mechanical continuity varies in the repeated simulation, see the variation in Figure \ref{Hydration Segmentation}.
For comparison purposes, the curves are then shifted to a common point time$_{cont}$, which represents the mechanical continuity time for each individual simulation.

As expected, the different moduli (Young and Shear) increase with the hydration process. As the chemical process occurs, bonds are generated between the aggregates, and the mechanical properties of the sample harden. Similar to the tendencies observed in \cite{Nguyen2024}, the evolution consists of two stages: an initial rapid phase followed by a slower final phase. This curve is correlated to the hydration evolution depicted in Figure \ref{Time Hydration}.
The microstructure remains weak at the initial mechanical continuity point (when there exists a continuous paste+source phase between the top and the bottom of the sample). As shown in Figure \ref{Microstructure Evolution Maps}, the bonds between the aggregates remain small. However, the bond sizes increase with the hydration, and it reinforces the sample. 

The results for $w/c=0.5$ obtained by the PF formulation appear closer to the estimated moduli from the real microstructure obtained by \cite{Thomas2009} than the results from \cite{Nguyen2024}.
Indeed, the microstructure computed with the \emph{CEMHYD3D} description is more discrete, and the different phases are less continuous than the ones obtained with a PF formulation. 
These localizations generate weak points in the \emph{CEMHYD3D} microstructure and weaken the sample.
Furthermore, it has been pointed out in Figure \ref{Time Mean Value Species} that the \emph{CEMHYD3D} formulation overestimates the volume fraction of the pore, weakening the microstructure. 
Because of these weakening origins, the moduli of the homogenized \emph{CEMHYD3D} microstructure appear much smaller than the intrinsic value of the cement paste (weakest phase after pore/water and dominant in the microstructure). 
The PF formulation appears more suitable than the \emph{CEMHYD3D} description.

The $w/c=0.3$ microstructure appears stiffer than the $w/c=0.5$ microstructure for the different moduli (Young and Shear). Indeed, Figure \ref{Microstructure Evolution Maps} emphasizes there are more source inclusions in the case $w/c=0.3$. These inclusions are stiffer, see Table \ref{Mechanical Parameter Phases}, and affect the mechanical behavior of the sample. 
As a reminder, the saturation limit has been defined as $w/c=0.42$. This limit is defined as the maximal source quantity with full hydration. If $w/c$ is smaller, there are remaining source inclusions (rigid phase) after the hydration. If $w/c$ is larger, there is not enough source to produce a saturation of the hydration.

Similar to the previous results, the PF simulations appear reproducible. The largest variations are located in the evolution of the Shear modulus.
Indeed, Figure \ref{Microstructure Evolution Maps} depicts the fact that the initial configuration (position of the source particles) is heterogeneous. 
The Shear modulus is more sensitive to such irregular aspects of the microstructure, the pores acting as weakness points.
In the same vein, it is relevant to emphasize the fact that the variance for the $w/c=0.5$ configuration is larger than for the $w/c=0.3$ configuration. 
Indeed, the irregular aspect of the initial configuration is even more visible for $w/c=0.5$ as there is less cement source/more pore space. 
The quality of the results would be better with a more homogeneous initial configuration, considering another algorithm for the generation.

\begin{figure}[ht]
    \centering
    \includegraphics[width=0.7\linewidth]{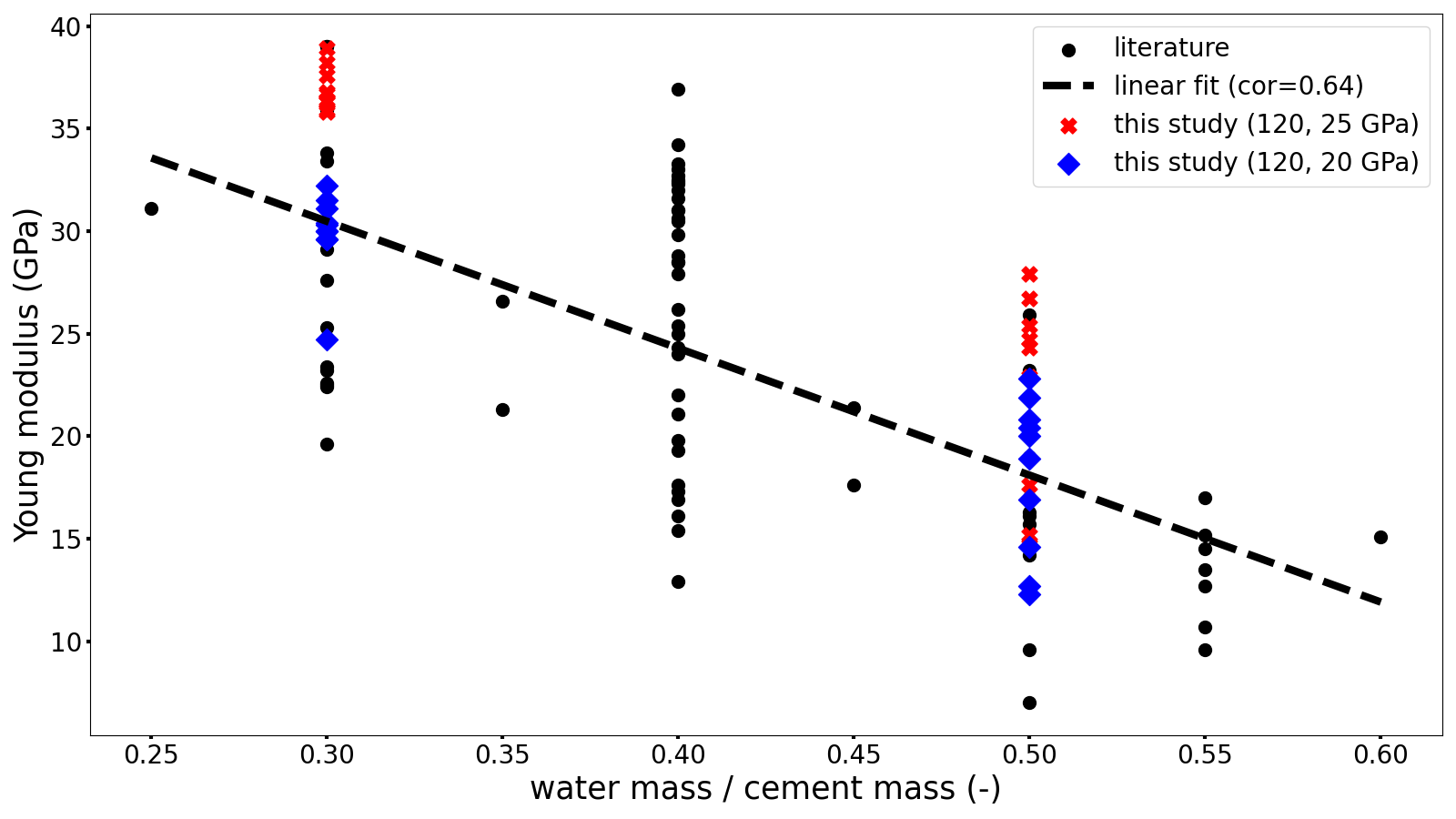}
    \caption{Dataset of the Young modulus of the hydrated concrete material for distinct $w/c$ ratios. Different computation homogenization has been conducted from the microstructure evolution depicted herein, considering ($Y_{source}$, $Y_{paste}$).}
    \label{wc Young Figure}
\end{figure}

The results obtained for the Young modulus of the hydrated sample are compared with the values available in the literature for distinct $w/c$ values in Figure \ref{wc Young Figure}. 
It is important to point out that these data have been obtained with a PF formulation + computational homogenization \cite{Petersen2018}, a CA description + computational homogenization \cite{Nguyen2024, Haecker2005, Qian2012, Hou2019, Li2022, Yan2022, Yu2023}, a computational homogenization based on experimental observations \cite{Lukovic2015, Zhang2016, Zhang2019, Savija2020, Zhang2024, Zhang2025}, a several-steps homogenization model \cite{Hu2014, Constantinides2004}, or by experiments \cite{Nguyen2024, Maruyama2014, Constantinides2004}. 
Moreover, sample characteristics are not always consistent in this dataset. 
For instance, the species distribution, the preparation methods, or the intrinsic mechanical parameters for the distinct phases are not similar. The properties of the composite material appear to be widely impacted by these fluctuations. 

First of all, it appears that, at a given $w/c$, a large distribution of the Young modulus of the hydrated sample is available in the literature. 
This parameter is deeply affected by the prediction method (modelization or experimental). 
In the case of the modelization, the homogenized Young modulus depends on the intrinsic value considered for the different phases and on the quality of the microstructure estimation.  
For instance, it has been emphasized in Figure \ref{Time Mean Value Species} that the CA approach tends to overestimate the pore space (weak phase) during the hydration process.
Then, it is not surprising to obtain larger values employing similar intrinsic parameters with a PF formulation. Indeed, the approach employed herein performs better in the description of the hydrated microstructure, predicting less pore space. 
Moreover, the C-S-H material is a heterogeneous phase with, in particular, a variation in the packing density \cite{Ioannidou2014, Hu2014, Goyal2020}. This fluctuation has an impact on the local properties of the cement material \cite{Moutassem2016} and may generate some weak regions.
The Young modulus estimated through a numerical homogenization method after the PF prediction of the microstructure appears in the upper envelope of the available dataset. 
However, the quality of the PF prediction appears to increase by considering the lower boundary depicted in \cite{Nguyen2024} for the intrinsic Young modulus of the C-S-H phase ($Y_{paste}=20$ GPa).

Obviously, the prediction of the method presented herein would increase, considering a 3D microstructure, capturing the fingering growth pattern of the paste, employing a more realistic size distribution and shapes for the source particle, or by adding the supplementary phases and their physics. 


\section{Discussion}

The numerical laboratory approach, describing explicitly the microstructure, provides a suitable framework for investigating the hydration process of concrete materials.
It appears that the currently most employed formulation is based on discrete models, such as \emph{CEMHYD3D}.
However, the coefficients used are largely derived from semi-empirical methods. 
Moreover, the intrinsic nature of CA induces that the microstructure obtained is highly heterogeneous, and different concentrations of species at one location in the pore water can not be modeled.
Consequently, a physics-based description, the PF formulation, seems more adequate. For instance, this work emphasizes that the PF method reproduces the microstructure evolution driven by chemical processes more accurately than the CA model. 

In the same vein, it is worth comparing the framework depicted herein with the analytical homogenization methods available in the literature. 
The difference is the fact that a PF describes the microstructure explicitly, while the homogenization approach requires making assumptions on the shape, the proportion, and the location of the distinct phases. 
Even if the microstructure is described, it is not in an explicit manner.
It has been emphasized that the microstructure is pivotal for the estimation of the material behavior (mechanics, chemistry, or hydrology) \cite{Lindqwister2025, Lindqwister2025b, Slotte2020}, emphasizing the importance of capturing it accurately.
Moreover, the PF approach enables the couplings with other physics, such as the temperature, the stress, or the presence of a fluid transport, while such considerations remain difficult through constitutive relations for the hydration of the material, a key ingredient of the prediction by homogenization.

Nevertheless, the computational cost of the PF simulations remains huge; the resolution of a 2D simulation (microstructure evolution prediction and mechanical behavior interpolation, depicted herein) lasts approximately 10 hours on a regular computer (Intel(R) Core(TM) i5-10210U CPU @ 1.60GHz, 8 Go RAM). 
However, numerous methods may be employed to diminish the resources required, allowing the use of the PF description for 3D problems, considering multi-physics \cite{Sessim2022, Battas2025}. Future work should extend this model to 3D to more accurately capture the complex pore network connectivity and tortuosity, which are known to be critical for transport properties and may further refine the predicted percolation threshold and mechanical response.


\section{Conclusion}

This paper presents a Phase-Field formulation of the hydration of a concrete material, describing explicitly the microstructure and its evolution. In particular, the proposed physics-based description—featuring a revised free-energy potential and distinct equilibrium constants—is compared to experimental observations and to a semi-empirical Cellular Automata approach available in the literature. Considering two distinct boundary value problems ($w/c = 0.3$ and $0.5$), the Phase-Field formulation appears to reproduce the kinetics, phase distribution, and organization of the microstructure. It has also been shown that the Cellular Automata approach tends to overestimate pore space during hydration, which is not observed with the formulation depicted herein.

Once the microstructural evolution is estimated with the Phase-Field model, the resulting microstructures are used as inputs to a computational homogenization scheme to predict the evolution of the mechanical properties of the material. Mechanical performance appears to be captured with more precision through the Phase-Field formulation than through the Cellular Automata description. In particular, the overestimation of porosity inherent to Cellular Automata leads to an artificially weakened composite, whereas the Phase-Field model provides a more physically consistent estimate.

Although this work focuses on the mechanical response of the hydrating material, the framework depicted herein can readily be extended to other coupled physics. With appropriate constitutive laws, the same microstructural predictions could be used to assess thermal or hydraulic behavior, offering a more comprehensive multiphysics view of cementitious materials.

The Phase-Field description therefore provides a promising foundation for improving our understanding of hydration at the microscale and for exploring the influence of key parameters such as particle size distribution or chemical composition. Although further validation and application to a wider set of conditions are still needed, the ability to anticipate mechanical behavior directly from the evolution of the simulated microstructure suggests a useful complementary tool alongside traditional 28-day testing protocols.


\section{Code availability}

Some examples of scripts used are available on Github: \url{https://github.com/AlexSacMorane/PF_CementHydration}.


\section{Acknowledgements}

This research has been partially funded by the Fonds Spécial de Recherche (FSR) and by the Wallonia-Bruxelles Federation.
The work has also received funding from the NSF project CMMI-2042325.


\appendix

\section{Particle Size Distribution of the source}
\label{PSD source}

To ensure the model is representative of the experiments, it is important to reproduce most of the parameters available. 
Concerning the hydration problem, the particle size distribution of the source grains represents a parameter with a large impact on the hydration kinetics and the microstructure evolution obtained.

In the context of a numerical analog, the investigation pursued employed a particle size distribution from a database available \cite{Nguyen2021,NIST2005}, see Table \ref{PSD Hydration}. This Table presents the mass and number ratios of different radius bins (from $0.5\;\mu m$ 
to $36.5\;\mu m$). Between the limits of the bins, a uniform distribution is assumed for the radius. 
However, the PF formulation induces a minimal size for grains to remain stable.
Indeed, if the particle size is smaller than the minimal size ($\sim 2\times \delta$, where $\delta$ is the interface width), the maximum value of the phase variable is smaller than $0.5$, dissolution $\left(\eta \rightarrow 0\right)$ occurs due to the PF formulation. 
As emphasized in Table \ref{Parameter Simulation Microstructure}, the mesh size $\Delta$ is $0.2\;\mu m$, limiting the minimal radius to $2\times\delta=2\times6\Delta=2.4\;\mu m$. The particle size distribution is then modified to consider this minimal size restriction, see Table \ref{PSD Hydration}. 


\begin{longtable}{|l|c|c|c|}
    \hline 
    Radius & Mass (\%) & Number (\%) & Number (\%) \\
    ($\mu m$) & initial & initial & from $2.5\;\mu m$ \\
    \hline
     0.5 & 13.5 & 95.8 &       \\
     1.5 & 13.0 &  3.4 & not considered \\
     \cline{4-4}
     2.5 &  9.5 &  0.5 & 66.3 \\
     3.5 &  7.3 &  0.2 & 18.6 \\
     4.5 &  6.1 &  0.1 &  7.3 \\
     \cline{3-3}
     5.5 &  5.1 &      &  3.3 \\ 
     6.5 &  4.2 &      &  1.7 \\
     7.5 &  3.8 &      &  1.0 \\
     8.5 &  3.4 &      &  0.6 \\
     9.5 &  2.8 &      &  0.4 \\
    10.5 &  2.6 &      &  0.2 \\
    11.5 &  2.3 &      &  0.2 \\
    12.5 &  2.1 & not enough & 0.1 \\
    13.5 &  1.9 & ($<$ 0.1)  & 0.1 \\
    14.5 &  1.9 &      & 0.1 \\
    15.5 &  2.3 &      & 0.1 \\
    \cline{4-4}
    17.5 &  3.5 &      &     \\
    20.5 &  2.7 &      & not enough \\
    23.5 &  4.5 &      & ($<$ 0.1) \\
    30.5 &  3.6 &      &     \\
    36.5 &  4.0 &      &     \\
    \hline
    \caption{Particle size distribution for the cement CCL 133.}
    \label{PSD Hydration}
\end{longtable}

The conversion between the mass fraction and the number fraction has been obtained by considering that the shape of the particle is a sphere and by following the scheme described in Algorithm \ref{Scheme Mass to Number ratio}.

\begin{algorithm}[ht]
\caption{Convert a mass ratio $V_i$ to a number ratio $N_i$.}
\label{Scheme Mass to Number ratio}
\begin{algorithmic}
\State assume there is $n_0=1$ particle for the size $0$
\ForAll {size $i \in$ bins (except $i=0$)}
    \State compute the number of particle $n_i$ for the size $i$ \Comment{$\frac{n_i}{n_0}=\frac{V_i}{V_0}\left(\frac{R_0}{R_i}\right)^3$} 
\EndFor
\State Compute the total number of particles $n_t$ \Comment{$n_t=\sum\limits_i n_i$} 
\ForAll {size $i \in$ bins}
    \State compute the number ratio $N_i$ for the size $i$ \Comment{$N_i=\frac{n_i}{n_t}$} 
\EndFor
\end{algorithmic}
\end{algorithm}


\section{Sensitivity analysis of the coefficient $\alpha_{CSH}$}
\label{alpha_CSH sensitivity}

\begin{figure}[h]
    \centering
    a) \includegraphics[width=0.45\linewidth]{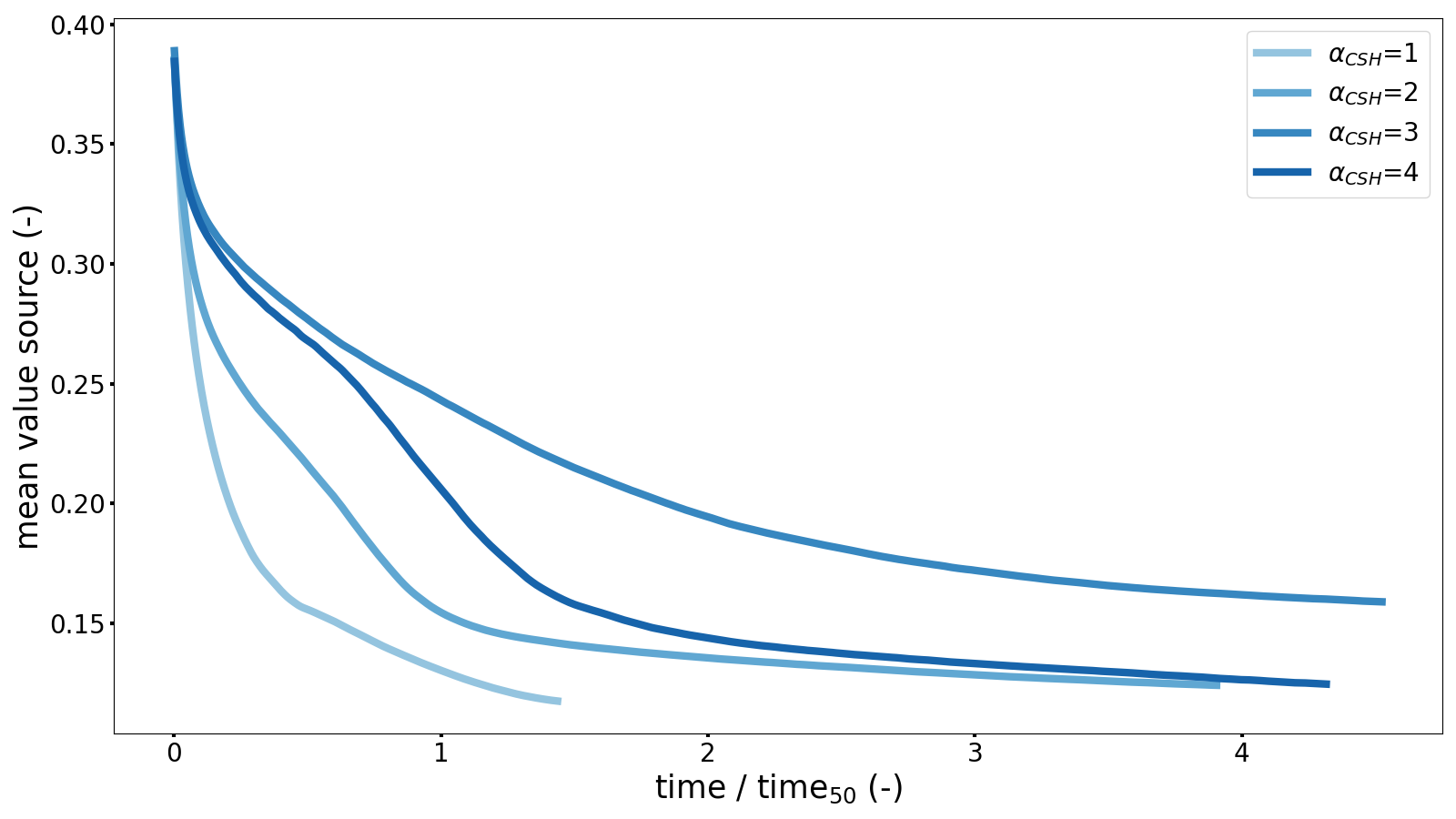}
    b) \includegraphics[width=0.45\linewidth]{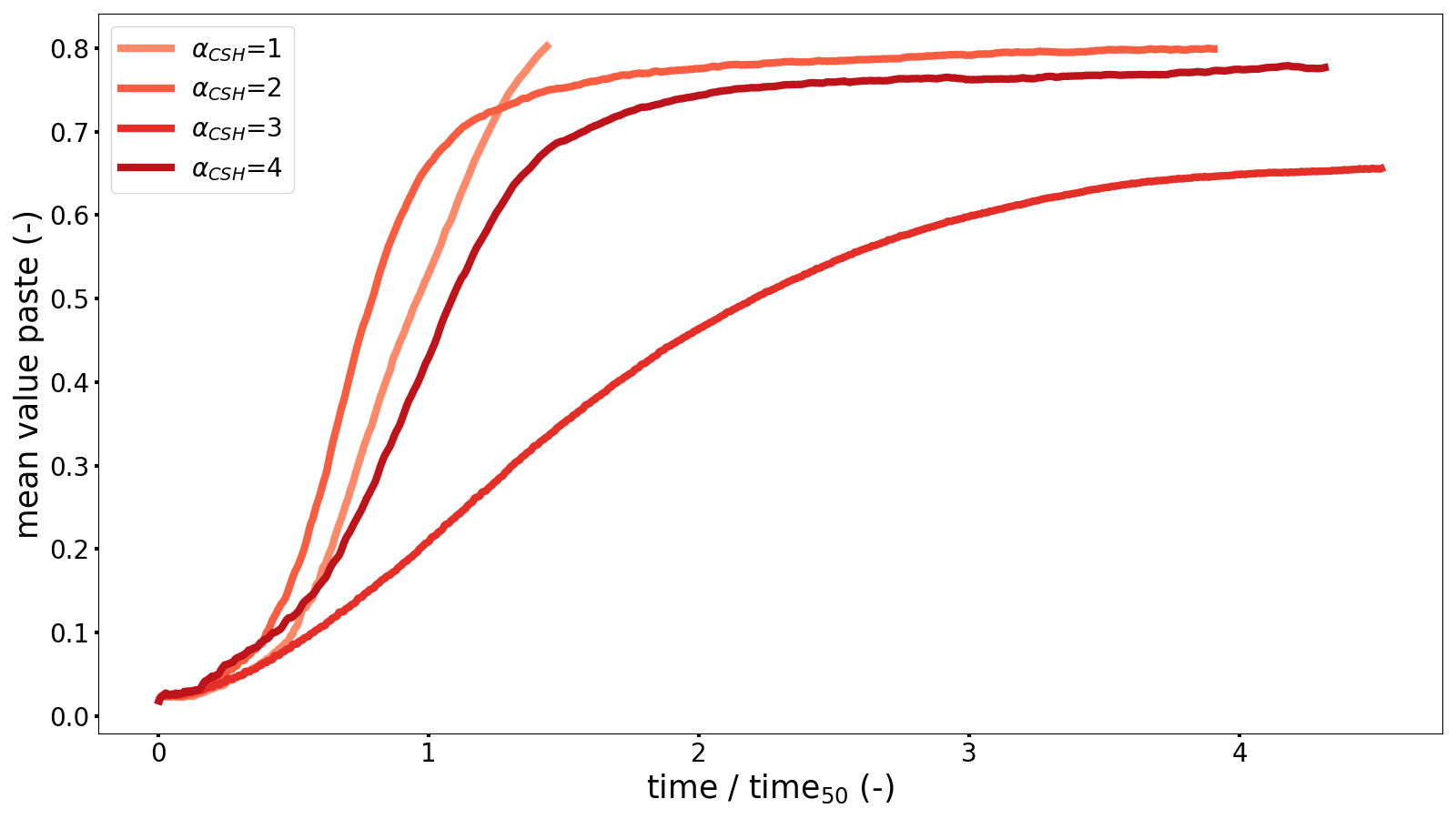}\\
    c) \includegraphics[width=0.45\linewidth]{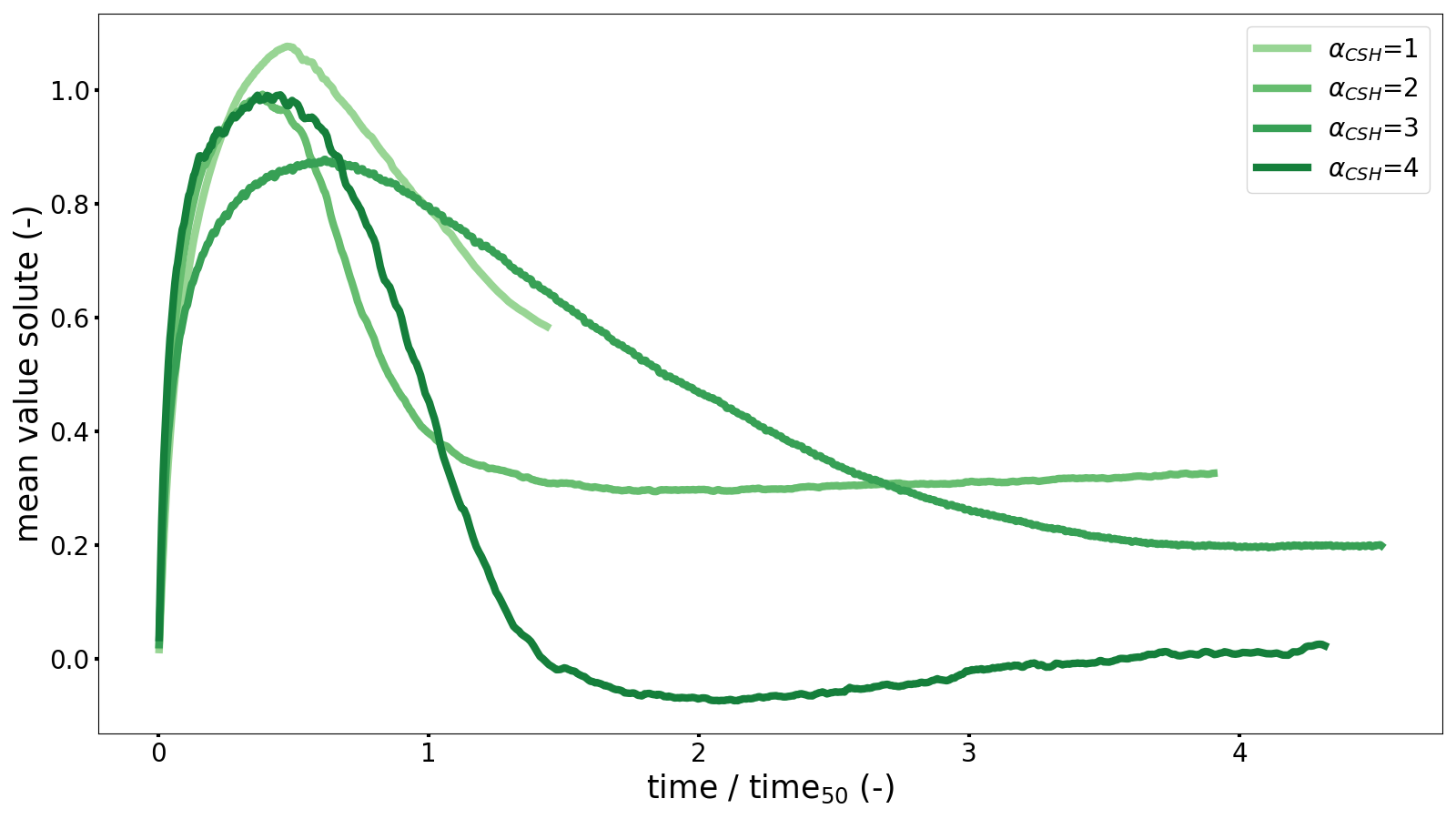}
    \caption{Evolution of the mean value of the variable a) $C3S$, b) $CSH$, and c) $c$ for different $a_{CSH}$ coefficients ($w/c=0.5)$.}
    \label{Time Mean Value Species a_CSH}
\end{figure}

As mentioned in Subsection \ref{Equations Section}, the conversion terms $\alpha_{CSH}/\alpha_{C3S}$ are correlated, but the values remain unknown.
In the absence of experimental data available, a sensitivity analysis is then conducted to determine the value of $\alpha_{CSH}$ (and $\alpha_{C3S}$ through Equation \ref{Mass Conservation Relation}). The results for the configuration $w/c=0.5$ are available in Figure \ref{Time Mean Value Species a_CSH}.

It appears that a small $\alpha_{CSH}$ induces the fact that the source particles are significantly dissolved before the precipitation of the paste, see the Figure \ref{Time Mean Value Species a_CSH}a), not in agreement with the experimental observations  \cite{Bullard2011, Scrivener2015}.
Indeed, the solute $c$ produced by the dissolution of the source is not sufficient to generate a supersaturation state.
On the contrary, a large  $\alpha_{CSH}$ instigates negative values for the solute concentration $c$, see the Figure \ref{Time Mean Value Species a_CSH}c). This state is explained by the large quantity of solute required for the paste precipitation.
Finally, the value $\alpha_ {CSH}=3$ is employed herein as it seems to verify the experimental observations (small source dissolution at the supersaturation state and positive value for the solute concentration).


\section{Definition of the normalized two-point correlation function}
\label{Definition Correlation}

In the goal of comparing microstructures, a normalized two-point correlation function has been developed \cite{Bentz2006}.
Considering an $M\times N$ image of the microstructure, the two-point correlation function is obtained through Equation \ref{Correlation Function x y}.

\begin{equation}
    S(x,y) = \sum\limits^{M-x}_{i=1}\sum^{N-y}_{j=1}\frac{I(i,j)I(x+i,\, y+j)}{(M-x)(N-y)}
    \label{Correlation Function x y}
\end{equation}

where $I(i,j)=1$ if the voxel located at $(i,j)$ represents the phase of interest, $=0$ otherwise. 
The values $x$ and $y$ are the pixel offsets investigated and can be converted into a radial distance $r$ \cite{Berryman1985}:

\begin{equation}
    S(r) = \frac{1}{2r+1}\sum\limits^{2r}_{l=0}S\left(r,\frac{\pi l}{4r}\right)
    \label{Correlation Function r}
\end{equation}

It appears that the value $S(0)$ represents the volume fraction of the investigated phase.
To compare the microstructure at different volume fractions, the two-point correlation function is normalized with Equation \ref{Correlation Function Normalized}. 

\begin{equation}
    N(r) = \frac{S(r)-S(0) S(0)}{S(0)-S(0)
    S(0)}
    \label{Correlation Function Normalized}
\end{equation}


\section{Correlation functions for repeated configurations}
\label{Correlation Repetition}

\begin{figure}[h]
    \centering
    a) \includegraphics[width=0.7\linewidth]{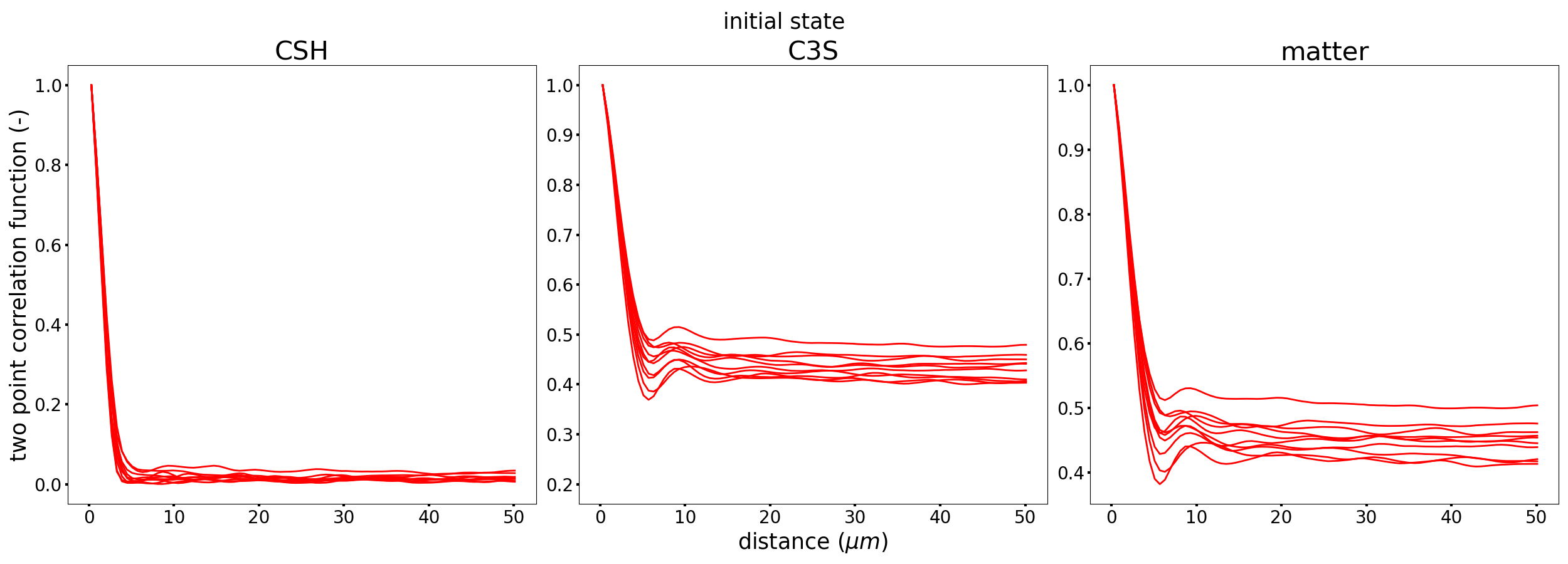}\\
    b) \includegraphics[width=0.7\linewidth]{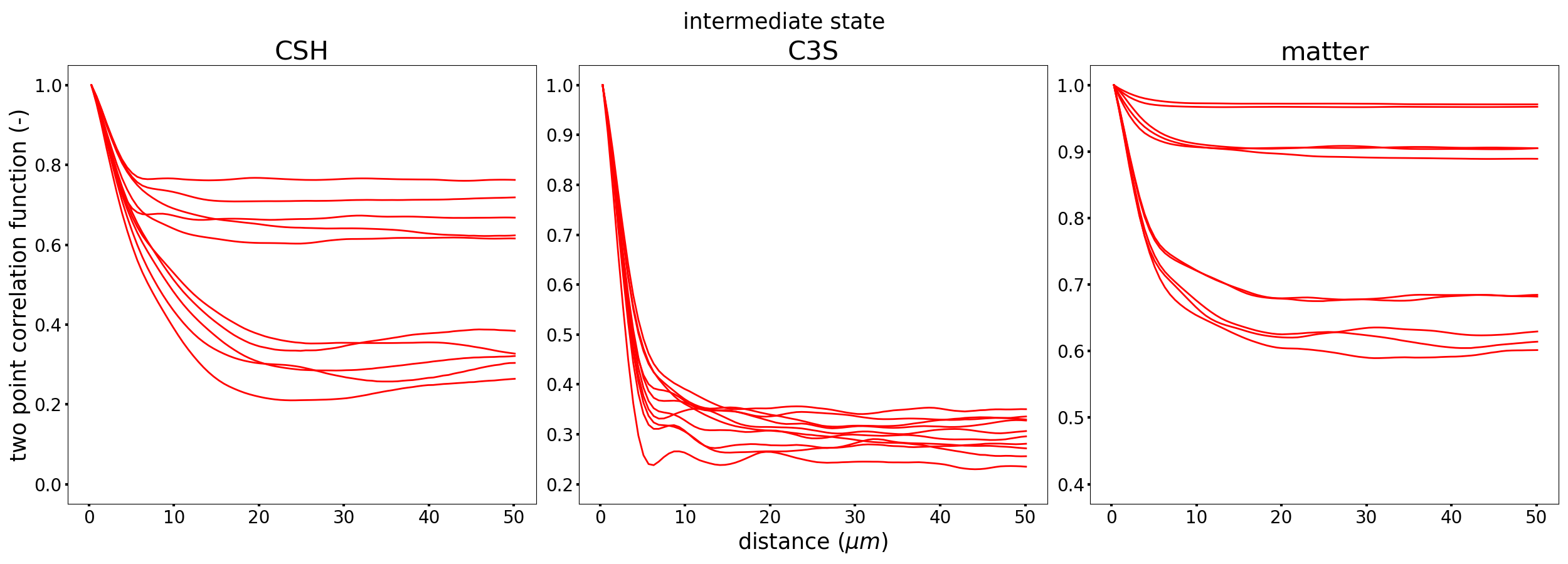}\\
    c) \includegraphics[width=0.7\linewidth]{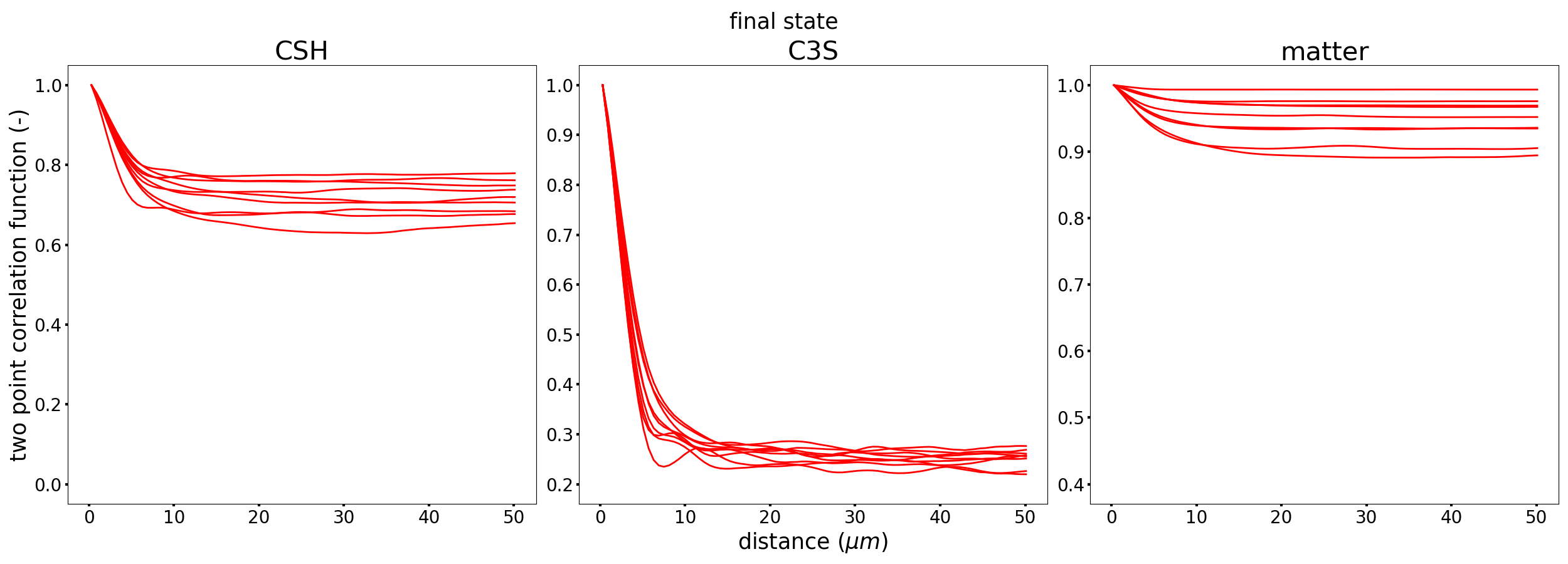}\\
    \caption{Evolution of the correlation function for the different phases (paste, source, paste+source) in the case w/c = 0.3 for $t/t_{50}=$ a) 0.04, b) 0.41, and c) 0.83. The configuration has been repeated ten times.}
    \label{Time Correlation Repetition}
\end{figure}

The two-point correlation function is a common tool to provide insights into the microstructure \cite{Bentz2006}. For instance, Figure \ref{Hydration Correlation} depicts the evolution of the functions during the hydration process, considering an individual simulation.
However, the repeatability is not emphasized, for clarity reasons.
This Appendix and Figure \ref{Time Correlation Repetition} investigate this aspect in the case w/c = 0.3.

It appears in Figure \ref{Time Correlation Repetition}a) and c) that the functions are overlaid for the initial and final states. The initial state of the configuration is unsurprisingly repeatable, while it is comforting to verify that the final state is similar in the different simulations. 
Concerning the intermediate state depicted in Figure \ref{Time Correlation Repetition}b), the range of the functions is narrow. 
In particular, two groups of curves appear. 
Indeed, Figure \ref{Time Hydration} emphasizes the variation of the hydration kinetics, in particular for the intermediate state ($t/t_{50}=0.41 \,\sim h=30\%$). 
Some simulations depict faster kinetics compared to the others, explaining the two groups.
The prediction of the microstructure evolution appears reproducible. 


\bibliography{bibliography}
\bibliographystyle{ieeetr}

\end{document}